\documentstyle[preprint,tighten,aps,epsfig]{revtex} 
\begin{document}   
\preprint{TAUP-2586-99, MKPH-T-01-07}  
\draft 
\tighten 
\title{Pion Generalized Dipole Polarizabilities by Virtual Compton 
Scattering $\pi{e}\rightarrow\pi{e}\gamma$} 
\author{ 
C. Unkmeir,$^1$  
A. Ocherashvili,$^2$ 
T. Fuchs,$^1$   
M. A. Moinester,$^2$ 
and  
S. Scherer$^1$} 
\address{ 
$^1$ Institut f\"ur Kernphysik, Johannes Gutenberg-Universit\"at, 
J.\ J.\ Becher-Weg 45, D-55099 Mainz, Germany\\ 
$^2$ School of Physics and Astronomy, R. and B. Sackler Faculty of
Exact Sciences,\\ 
Tel Aviv  University, 69978 Tel Aviv, Israel} 
\date{July 3, 2001} 
\maketitle 
\begin{abstract} 
      We present a calculation of the cross section and the event 
   generator of the reaction $\pi e\to \pi e \gamma$.  
      This reaction is sensitive to the pion generalized dipole  
   polarizabilities, namely, the longitudinal electric $\alpha_L(q^2)$, 
   the transverse electric $\alpha_T(q^2)$, and the magnetic  
   $\beta(q^2)$ which, in the real-photon limit, reduce to the ordinary 
   electric and magnetic polarizabilities $\bar{\alpha}$ and 
   $\bar{\beta}$, respectively.  
      The calculation of the cross section is done in the framework of chiral 
   perturbation theory at ${\cal O}(p^4)$.  
      A pion VCS event generator has been written which is ready for  
   implementation in GEANT simulation codes or for independent use. 
\end{abstract} 
 

\section{Introduction} 
\begin{figure}[ht]
\vspace{0.5em} 
\epsfxsize=10cm
\centerline{\epsffile{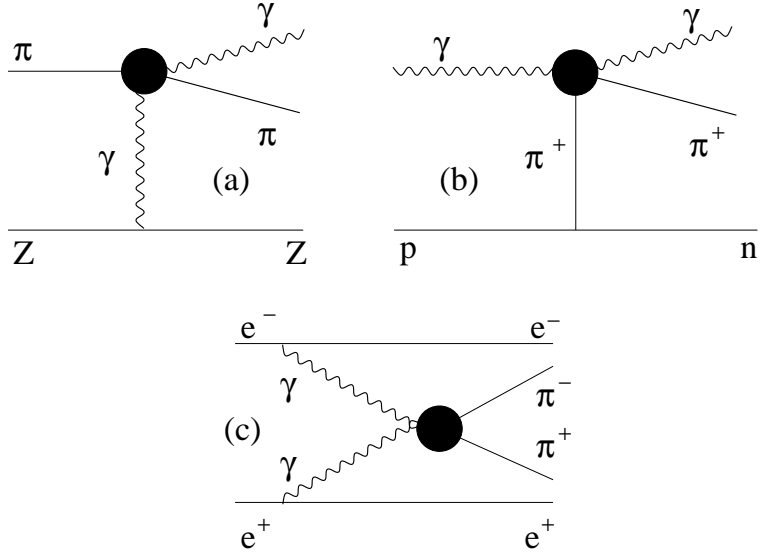}}
\vspace{1em} 
\caption[]{Three pion Compton scattering reactions: 
             (a) $\pi^- Z\rightarrow\pi^- Z \gamma$,  
             (b) $\gamma p\rightarrow\gamma\pi^+ n$, 
             (c) $e^+ e^-\rightarrow{e^+ e^-} \pi^+ \pi^-$.   
\label{fig:polpos}} 
\end{figure}

   Pion electromagnetic polarizabilities are important dynamical 
properties, providing a window to the pion's internal structure 
(for an overview see, e.g., Refs.\ \cite{overview}).  
   In a classical picture, polarizabilities determine the global, 
i.e.\ integral, deformation response of a composite system in static,  
uniform external electric and magnetic fields.  
   As first shown in Refs.\ \cite{kbp},  
a determination of hadron polarizabilities  
is possible via $\gamma$-hadron Compton scattering.  
   So far, pion electric ($\bar{\alpha}$) and magnetic ($\bar{\beta}$) 
polarizabilities were investigated via $\gamma\pi\rightarrow\gamma\pi$ 
real Compton scattering (RCS).  
   Since pion targets are unavailable, pion RCS was done using different  
artifices, such as high-energy pion-nucleus bremsstrahlung,  
$\pi^- Z\rightarrow\pi^- Z \gamma$ \cite{Antipov}, 
radiative pion photoproduction off the proton 
$\gamma p\rightarrow\gamma \pi^+ n$ \cite{aibergenov}, 
and the crossed-channel two-photon reaction  
$\gamma\gamma\rightarrow\pi\pi$ as embedded in the 
$e^- e^+\rightarrow{e^- e^+}\pi^+\pi^-$ process \cite{Boyer,Pole} 
(see Fig.\ \ref{fig:polpos}).  
   The values of $\bar{\alpha}$ 
extracted from these experiments are given in Table \ref{tab:mespol}.  
 
   As can be seen, the results scatter substantially and have large  
uncertainties, and therefore, more experimental efforts are evidently needed. 
   A new radiative pion photoproduction experiment at the Mainz   
Microtron MAMI has recently taken data \cite{A2}, another one has  
been proposed at Jefferson Lab \cite{norum99}, an experiment  
investigating the inverse reaction $\pi^-p\to n\gamma\gamma$ has been proposed
at TRIUMF \cite{gorringe}, a new 
$\gamma\gamma\rightarrow\pi\pi$ experiment was proposed 
at Frascati \cite {gg}, and new Primakoff (bremsstrahlung) 
experiments were proposed at CERN COMPASS \cite {tm}.
 
   Electromagnetic studies with virtual photons are advantageous, since 
the energy and the modulus of the three-momentum of the virtual photon  
can be varied independently.  
   Furthermore, one also gets access to longitudinal degrees of freedom 
which cannot be tested by real photons. 
   In the pion VCS reaction  $\gamma^*\pi\rightarrow\gamma\pi$  
with a virtual space-like initial-state photon and a real 
final-state photon, the so-called generalized polarizabilities (GPs) 
can be measured in the space-like region. 
   Originally, such quantities were introduced in the context of nuclei 
\cite{arenhoevel74} and have recently been reinvestigated for the 
nucleon in Ref.\ \cite{guichon95}. 
   The analysis of the nucleon GPs is based on a multipole expansion 
in the center-of-mass frame below the pion production threshold. 
   The first experimental results for generalized polarizabilities  
of the proton have been reported by the VCS collaboration at MAMI  
\cite{Roche_2000} and are in good agreement with the 
predictions of chiral perturbation theory \cite{hemmert97}. 
 
   Recently, another approach to the analysis of the VCS tensor has been  
performed in a completely covariant framework \cite{unkmeir00,lvov01}. 
   In particular, the reaction $\gamma^\ast\pi\to\gamma\pi$ is described 
in terms of three Lorentz-invariant functions, each depending on three scalar 
variables. 
   In the limit of a soft final real photon, $q'\to 0$, the 
structure-dependent 
response is encoded in {\em three} generalized dipole polarizabilities 
$\alpha_L(q^2)$, $\alpha_T(q^2)$, and $\beta(q^2)$ \cite{lvov01}. 
   The Fourier transforms of these functions are related to the  
electric polarization and magnetization induced by soft external  
fields \cite{lvov01}. 
   In other words, the VCS generalized dipole polarizabilities allow one 
to access the local polarization response, whereas the RCS polarizabilities 
$\bar{\alpha}=\alpha_L(0)=\alpha_T(0)$ and $\bar{\beta}=\beta(0)$ only 
provide the integrated, i.e.\ global, response. 
   Furthermore, {\em two} electric polarizabilities are needed to  
fully reconstruct the electric polarization, the longitudinal electric  
polarizability being related to the divergence of the induced electric  
polarization, and the transverse electric polarizability describing rotational
displacements of charges which do not result in a modification of the 
charge density (such as, e.g., a rotating spherical charge distribution). 
   On the other hand, due to the transverse character of the magnetic 
induction, only one generalized magnetic dipole polarizability appears.

   Motivated by the success of the nucleon VCS experiment \cite{Roche_2000}, 
by studying the reaction  
\begin{equation} 
\label{react} 
\pi e\rightarrow\pi e\gamma, 
\end{equation} 
   we attempt to get a hold of the pion VCS matrix element 
$\gamma^\ast(q)+\pi(p_i)\to\gamma(q')+\pi(p_f)$,  
which can then be utilized as a new experiment for a pion polarizability  
measurement.  
 
   For the numerical analysis of the above reaction, we make 
use of the results obtained in the framework of chiral perturbation 
theory at ${\cal O}(p^4)$ \cite{unkmeir00}. 
   At this order in the momentum expansion, the generalized dipole 
polarizabilities display a degeneracy,  
$\beta(q^2)=-\alpha_L(q^2)=-\alpha_T(q^2)$ 
which will be lifted at ${\cal O}(p^6)$ and ${\cal O}(p^8)$, respectively. 
 
   Our work is organized as follows. 
   In Sec.\ II, we briefly discuss the predictions for the generalized 
dipole polarizabilities as obtained in the framework of ChPT at 
${\cal O}(p^4)$ \cite{unkmeir00}. 
   We provide a physical interpretation of the longitudinal  
generalized electric dipole polarizability in terms of the induced 
polarization 
charge\cite{lvov01}. 
   Section III briefly describes the kinematics and notation used. 
   Section IV contains a discussion of the various pieces entering 
the cross section, whereas some general features of the cross section are  
discussed in Sec.\ V. 
   Section VI introduces the ingredients of the event generator.
   In Sec.\ VII we discuss the sensitivity of the cross section to 
the polarizabilities.  
   Section VIII contains the results for the cross section obtained after  
Monte Carlo integration.  
   The present experimental status is summarized in Sec.\ IX. 
   In Sec.\ X we summarize our results and draw some conclusions. 
   The analytical results for the squared invariant amplitudes are 
given in the Appendix. 
 
\section{Virtual Compton Scattering and generalized dipole polarizabilities} 
\label{sec:vcsi} 
\begin{figure}[ht]
\epsfxsize=8cm
\centerline{\epsffile{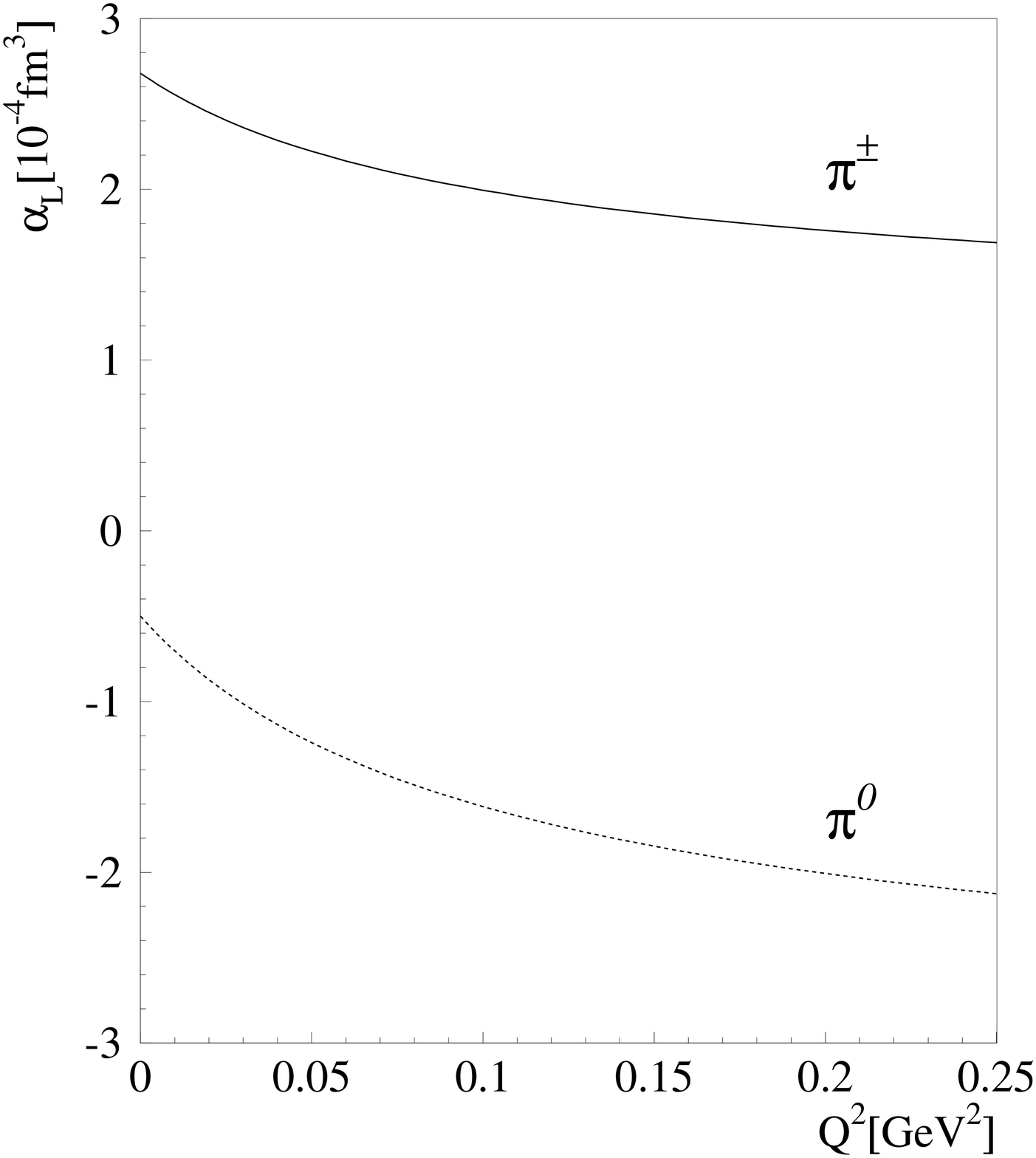}}
\vspace{1em}
\caption[]{${\cal O}(p^4)$ prediction for the generalized dipole 
     polarizabilities $\alpha_L(-Q^2)$ of the charged pion (solid curve) 
     and the neutral pion (dashed curve) as functions 
     of $Q^2$ [see Eqs.\ (\ref{alphapp1}) and (\ref{alphap02})].
    \label{fig:beide} }
\end{figure}

   A new frontier for polarizability studies is the subject of VCS 
\cite{Roche_2000,holstein,scherer}.  
   There are two different ways of studying this reaction in different 
kinematical regimes regarding the virtual photon.  
   In the first case, the initial photon is real and the final virtual 
photon fragments into a Dalitz pair \cite{lvov}.  
   This corresponds to positive $q'^2\geq 4m_e^2$ and will not be discussed  
here.  
   Rather, we concentrate on the second case, where the initial virtual photon 
is produced in a particle scattering process, thus leading to $q^2<0$,  
and then scatters from a target to a real final-state photon.  
   This process probes pion structure via the generalized dipole  
polarizabilities $\alpha_L(q^2)$, $\alpha_T(q^2)$, and $\beta_L(q^2)$ 
which in the real-photon limit, $q^2\to 0$, reduce to the RCS polarizabilities 
$\alpha_L(0)=\alpha_T(0)=\bar{\alpha}$ and $\beta(0)=\bar{\beta}$ 
\cite{unkmeir00,lvov01}. 
   The predictions in the framework of ChPT at ${\cal O}(p^4)$  
read \cite{unkmeir00} 
(see Fig.\ \ref{fig:beide}) 
\begin{eqnarray} 
\label{alphapp1} 
\alpha^{\pi^\pm}_L(q^2)&=& 
\alpha^{\pi^\pm}_T(q^2)= 
-\beta^{\pi^\pm}(q^2)\nonumber\\ 
&=&\frac{e^2}{8\pi{m_{\pi}}} 
\left[ 
\frac{4(2l^r_5-l^r_6)}{F^2_{\pi}}- 
\frac{q^2}{m^2_{\pi}}\frac{1}{(4\pi{F_{\pi}})^2}J^{(0)'} 
\left(\frac{q^2}{m^2_{\pi}}\right) 
\right]\equiv\bar{\alpha}-f(q^2),\\ 
\label{alphap02} 
\alpha^{\pi^0}_L(q^2)&=& 
\alpha^{\pi^0}_T(q^2)= 
-\beta^{\pi^0}(q^2)\nonumber\\ 
&=&\frac{e^2}{4\pi m_\pi}\frac{1}{(4\pi F_\pi)^2} 
\left(1-\frac{q^2}{m^2_\pi}\right) 
J^{(0)'}\left(\frac{q^2}{m^2_\pi}\right), 
\end{eqnarray} 
where 
\begin{equation} 
\label{j0} 
J^{(0)'}(x)=\frac{1}{x} 
\left[ 
1-\frac{2}{x\sigma(x)} 
\ln{\left(\frac{\sigma(x)-1}{\sigma(x)+1}\right)}\right],\quad 
\sigma{(x)}=\sqrt{1-\frac{4}{x}},\quad 
x\le{0}. 
\end{equation} 
   The predictions for the charged pion RCS polarizabilities result from 
an old current-algebra low-energy theorem \cite{terentev72} 
\begin{equation} 
\label{alphapp11} 
\bar{\alpha}_{\pi^\pm}=-\bar{\beta}_{\pi^\pm} 
=\frac{e^2}{4\pi{m_{\pi}}}\frac{2(2l^r_5-l^r_6)}{F^2_{\pi}} 
=(2.68\pm{0.42})\times{10^{-43}}\mbox{cm}^3, 
\end{equation} 
   where the low-energy constants are fixed via the radiative 
pion decay $\pi^+\to e^+\nu_e\gamma$. 
   Corrections at ${\cal O}(p^6)$ were 
shown to be 12\% and 24\% of the ${\cal O}(p^4)$ values for  
$\bar{\alpha}$ and $\bar{\beta}$, respectively \cite{buergi96}.  
   The $q^2$ evolution of Eq.\ (\ref{alphapp1}) [and also of  
(\ref{alphap02})] is completely determined in terms of the pion mass 
$m_\pi$ and the pion-decay constant $F_\pi=92.4$ MeV.

   The generalized dipole polarizability $\alpha_L(q^2)$ can be interpreted 
in terms of the influence of a static, uniform electric field $\vec{E}$ on  
the pion charge form factor \cite{lvov01}.  
   In the rest of this section we will not make use of a covariant notation. 
Breit frame kinematics has to be applied in the transition to the  
relativistic case. 
   The presence of a static uniform electric field $\vec{E}$ will induce 
an electric polarization, the divergence of which is defined as the 
polarization charge density, 
\begin{equation} 
\label{drho} 
\delta\rho(\vec{x})=-\vec{\nabla}\cdot \vec{P}(\vec{x}). 
\end{equation} 
   The potential created by the original charge distribution in combination  
with the induced electric polarization will look like the one which is due 
to an effective charge density: 
\begin{equation} 
\label{chd} 
\rho_{\rm eff}(\vec{x})=\rho_0(\vec{x})-\vec{\nabla}\cdot\vec{P}(\vec{x}),   
\end{equation} 
where $\rho_0(\vec{x})$ is the charge density without external 
electric field. 
   The form factor of this effective charge density is then given by 
\begin{equation} 
\label{formf} 
F(\vec{q})=\int d^3 x\exp{(-i\vec{q}\cdot\vec{x})}\rho_{\rm eff}(\vec{x}) 
= F_0(\vec{q})+\Delta{F}(\vec{q}), 
\end{equation} 
where $F_0(\vec{q})$ is the form factor of the ``undistorted'' charge 
distribution, and 
\begin{equation} 
\label{dfq} 
\Delta{F}(\vec{q})=4\pi{i}\vec{q}\cdot\vec{E}\bar{\alpha}_L(\vec{q}) 
\end{equation} 
is the lowest-order change in the form factor due to the external electric  
field.  
 
   In the above discussion we have concentrated on the polarization charge 
density, i.e., the divergence of the electric polarization, or, in terms of  
its Fourier transform, the longitudinal contribution, only.  
   A full recovery of the polarization requires the transverse piece as well. 
   For details, the interested reader is referred to the comprehensive 
discussion of Ref.\ \cite{lvov01}.

\section{VCS kinematics and differential cross section} 
\label{sec:vcsk} 
   One possibility of studying the pion VCS matrix element would consist of 
investigating the electron scattering reaction  
$ep\rightarrow{e'}\gamma \pi^+ n$ 
as an extension of the RCS experiment 
$\gamma{p}\rightarrow\gamma\pi^+ n$ 
\cite{aibergenov,A2}. 
   In such a reaction, the VCS contribution is hidden as a subprocess 
in terms of a pion-pole diagram in the $t$ channel. 
   However, the initial pion is off its mass shell and thus the extraction 
involves an extrapolation to the unphysical region with all its ambiguities. 
 
   In the present work, we consider the pion VCS reaction  
$\pi(p_i)+e(k)\to\pi(p_f)+e(k')+\gamma(q',\epsilon')$ which, in principle,  
has the advantage of a direct access to the VCS matrix element without  
any extrapolation to the pion pole.   
   Here the pion four-momenta in the initial and final states are  
denoted by $p_i^\mu=(E_i,\vec{p}_i)$ and $p_f^\mu=(E_f,\vec{p}_f)$, 
respectively.  
   Similarly, the electron four-momentum reads $k^\mu=(E_k,\vec{k})$  
[$k'^\mu=(E_{k'},\vec{k}')$] and the final-state photon is 
characterized by its four-momentum $q'^\mu=(E_{q'},\vec{q}\,')$ 
and polarization vector $\epsilon'^\mu=(\epsilon'^0,\vec{\epsilon}\,')$. 
   The starting point of our discussion is the differential cross section in  
the convention of Ref.\ \cite{bdrel}:  
\begin{eqnarray} 
\label{bdcross} 
d \sigma&=&\frac{m_e^2}{8E_fE_{k'}E_{q'}} 
\frac{1}{\sqrt{(p_i\cdot k)^2-m_{\pi}^2m_e^2}}\frac{1}{(2\pi)^5} 
\nonumber\\&&\times  
\overline{|{\cal M}|^2} \delta^4(p_i+k-p_f-k'-q')d^3p_fd^3k'd^3q', 
\end{eqnarray} 
   where the invariant amplitude $\cal{M}$ contains the complete information  
on the dynamics of the process.  
   As usual, the quantity $\overline{|{\cal M}|^2}$ indicates an average 
over the initial polarizations and a summation over the final ones. 
 
\begin{figure}[ht]
\epsfxsize=8cm
\centerline{\epsffile{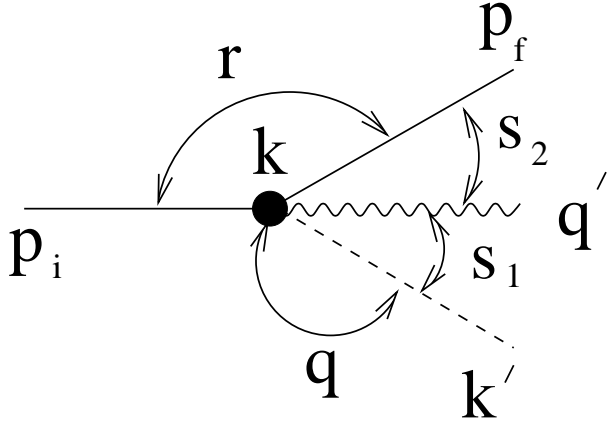}}
\vspace{1em}
\caption[]{Kinematics of $\pi(p_i)+e(k)\to\pi(p_f)+e(k')+\gamma(q')$ 
       in the laboratory (lab) frame.\label{fig:kin}} 
\end{figure}

    The cross section is more conveniently expressed in terms of the five  
independent invariant variables (see Fig.\ \ref{fig:kin}) 
\begin{eqnarray} 
\label{ivar} 
s&=&(p_i+k)^2, \nonumber \\ 
s_1&=&(k'+q')^2, \nonumber \\ 
s_2&=&(p_f+q')^2, \\ 
t_1&=&q^2=(k-k')^2, \nonumber\\ 
t_2&=&r^2=(p_i-p_f)^2. \nonumber  
\end{eqnarray} 
   This introduces a Gram determinant \cite{kinematic}  
\begin{equation} 
\label{eq:delta} 
\Delta_4=\frac{1}{16} 
\left| 
\begin{array}{cccc} 
2m^2_{\pi}         & s-m^2_{\pi}-m^2_e   & 2m^2_{\pi}-r^2      &  
s-s_2+q^2-m^2_e \\ 
s-m^2_{\pi}-m^2_e  & 2m^2_e              & s-s_1+r^2-m^2_{\pi} & 2m^2_e-q^2 \\ 
2m^2_{\pi}-r^2     & s-s_1+r^2-m^2_{\pi} & 2m^2_{\pi}          & s-s_1-s_2\\ 
s-s_2+q^2-m^2_e    & 2m^2_e-q^2          & s-s_1-s_2           & 2m^2_e 
\end{array} 
\right| 
\end{equation} 
into the expression for the fourfold differential cross section 
\begin{equation} 
\label{eq:xsection} 
\frac{d\sigma}{ds_1ds_2dq^2dr^2}= 
\frac{1}{{(2\pi)}^5} 
\frac{2m_e^2}{\lambda{(s,m_e^2,m_{\pi}^2)}} 
\frac{\pi}{16} 
\frac{1}{{(-\Delta_4)}^{1/2}} 
\overline{|{\cal M}|^2}, 
\end{equation} 
where  
\begin{equation} 
\label{eq:lambda} 
\lambda(s,m_e^2,m_{\pi}^2)=s^2-2(m_e^2+m_{\pi}^2)s+(m_e^2-m_{\pi}^2)^2. 
\end{equation} 
   The physical region is determined by the condition $\Delta_4<0$. 
   In an actual experimental realization, the energy of the pion beam 
is fixed such that $s=m_\pi^2+m_e^2+2E_i m_e$ in terms of laboratory 
variables. 
   In other words, the differential cross section of  
Eq.\ (\ref{eq:xsection}) then only depends on four variables. 
 
\section{Calculation of the squared invariant amplitude} 
\begin{figure}[ht]
\vspace{1em}
\epsfxsize=8cm
\centerline{\epsffile{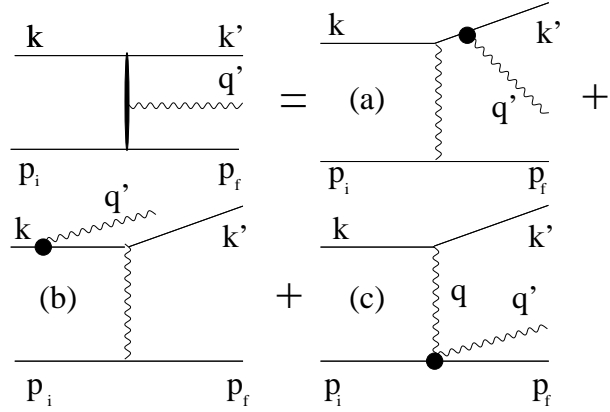}}
\vspace{1em}
\caption[]{The reaction $\pi^\pm(p_i)+e^-(k)\to \pi^\pm(p_f)+e^-(k') 
             +\gamma(q')$ to lowest order in the electromagnetic  
             coupling: Bethe-Heitler diagrams (a) and (b),  
             VCS diagram (c). The four-momentum of the virtual photon 
             emitted by the pion in the BH process is $r=p_i-p_f$. 
             The four-momentum of the virtual photon absorbed in the  
             pion VCS diagram is $q=k-k'$. Due to four-momentum conservation 
             $r=q'-q$.\label{fig:fdiag}} 
\end{figure}

   At lowest order in the electromagnetic coupling, the total invariant  
amplitude of the reaction  
$\pi^\pm(p_i)+e^-(k)\to \pi^\pm(p_f)+e^-(k')+\gamma(q')$  
can be divided into Bethe-Heitler (BH) and VCS contributions,  
${\cal M}={\cal M}_{\rm BH}+{\cal M}_{\rm VCS}$ (see Fig.\ \ref{fig:fdiag}). 
   In the case of the BH diagrams of (a) and (b), the 
real photon is emitted by the final and initial electrons, respectively. 
   The hadronic part of the corresponding amplitude is completely determined  
in terms of the pion electromagnetic form factor. 
   We denote the space-like four-momentum of the virtual photon emitted by 
the pion transition current in the BH process by $r=p_i-p_f=q'+k'-k$. 
   Due to the hermiticity of the electromagnetic current operator, the  
electromagnetic form factor is real in the space-like region \cite{scheck}. 
   Applying the Feynman rules of QED, it is straightforward to evaluate 
the BH amplitude.

   The pion VCS amplitude is part of the VCS diagram (c) of  
Fig.\ \ref{fig:fdiag}. 
   It is this contribution which contains the structure information in terms 
of the generalized dipole polarizabilities (and, of course, also of  
higher-order contributions). 
    
   According to Low's theorem \cite{low}, the irregular terms of the 
amplitudes ${\cal M}_{\rm BH}$ and ${\cal M}_{\rm VCS}$ behave as $1/E_{q'}$, 
where the singularities originate from the emission of the (soft) 
final photon at external lines. 
   Gauge invariance then fixes terms of ${\cal O}(E_{q'}^0)$ 
which do not depend on the direction $\hat{q}\,'$ 
(for a detailed discussion, see Refs.\ \cite{scherer96,fearing98}). 
   Model-dependent terms in ${\cal M}_{\rm VCS}$ are at least of  
order $E_{q'}$. 
   In this context, the VCS amplitude is divided into generalized 
Born terms ${\cal M}_{\rm Born}$, containing ground-state properties in terms  
of the electromagnetic form factor, and a residual non-Born contribution  
${\cal M}_{\rm NB}$, describing model-dependent internal  
structure (see Refs.\ \cite{unkmeir00,lvov01,fearing98}). 
   In particular, the separation is performed such that both pieces  
are separately gauge invariant and the generalized Born terms contain 
all soft-photon singularities. 
   The non-Born part is of ${\cal O}(E_{q'})$. 
    
   When squaring the total amplitude, the irregular terms of  
$\overline{|{\cal M}|^2}$ are of order $E_{q'}^{-2}$ and $E_{q'}^{-1}$.  
   Such terms entirely originate from the BH part and the generalized 
Born terms of the VCS part. 
   In particular, these contributions are completely known and can 
be expressed in terms of the electromagnetic form factor. 
   Recall that the space-like virtual-photon four-momenta are $r$ and $q$ in 
the Bethe-Heitler and VCS contributions, respectively, 
and only in the limit $q'\to 0$ they coincide. 
 
   The spin-averaged squared invariant amplitude can be 
decomposed into three parts, a pure VCS part, a pure BH part, and the 
interference term between BH and VCS:    
\begin{eqnarray} 
\label{mbhmvcs} 
\overline{|{\cal M}|^2}&=&\frac{1}{2}\sum_{s,s', \lambda'} 
({\cal M}_{\rm VCS}+{\cal M}_{\rm BH}) 
({\cal M}_{\rm VCS}^\ast+{\cal M}_{\rm BH}^\ast) 
\nonumber\\ 
&=& \overline{|{\cal M}_{\rm BH}|^2} 
+\overline{|{\cal M}_{\rm VCS}|^2} 
+\overline{{\cal M}_{\rm VCS}{\cal M}_{\rm BH}^\ast 
+{\cal M}_{\rm BH}{\cal M}_{\rm VCS}^\ast}. 
\end{eqnarray} 
The Lorentz invariant matrix element for the BH amplitude reads 
\begin{eqnarray} 
{\cal M}_{\rm BH}&=&\pm\frac{i e^3}{r^2}\epsilon^{\prime\ast\mu}(q',\lambda') 
\left[(p_f+p_i)^\nu F\left(r^2\right)\right] 
L_{\mu\nu}^{\rm BH},   
\end{eqnarray} 
  where the upper (lower) sign applies in the case of a  
$\pi^+$ ($\pi^-$) beam. 
   Here, $L_{\mu\nu}^{\rm BH}$ is nothing else but the second-rank electron  
Compton tensor, 
\begin{eqnarray} 
\label{lmunubh} 
L_{\mu\nu}^{\rm BH}&=&\bar{u}\left(k',s'\right) 
\left[\gamma_\mu\frac{1}{\gamma \cdot(k^{\prime}+q^{\prime})-m_e}\gamma_\nu 
+\gamma_\nu\frac{1}{\gamma\cdot(k-q')-m_e}\gamma_\mu\right] 
u\left(k,s\right)\nonumber\\ 
&=&\bar{u}\left(k',s'\right)\left[ 
\frac{\gamma_\mu \gamma\cdot q'+2 k'_\mu}{2k'\cdot q'}\gamma_\nu 
+\gamma_\nu \frac{\gamma\cdot q'\gamma_\mu -2 k_\mu}{2k\cdot q'}\right] 
u(k,s). 
\end{eqnarray} 
   Equation (\ref{lmunubh}) nicely displays the origin of the  
singularity for $q'\to 0$ as being due to the internal electron lines  
of the electron Compton scattering $s$- and $u$-channel diagrams approaching 
the on-shell limit. 
   The squared matrix element decomposes into the product of a pure 
leptonic and a pure hadronic part, having the same overall sign for 
both $\pi^+$ and $\pi^-$, 
\begin{eqnarray} 
\label{mbhsquared} 
\overline{|{\cal M}_{\rm BH}|^2}&=&  
\frac{1}{2}\sum_{s,s',\lambda'}{\cal M}_{\rm BH}{\cal M}_{\rm BH}^\ast 
\nonumber\\ 
&=& \left[\frac{e^3 F(r^2)}{r^2}\right]^2 
\left(p_i+p_f\right)^\mu \left(p_i+p_f\right)^\nu 
\overline{l^{\rm BH}_{\mu\nu}},  
\end{eqnarray} 
with  
\begin{eqnarray} 
\label{lbh} 
\overline{l^{\rm BH}_{\mu\nu}}&=&\frac{1}{2}\sum_{s,s',\lambda'} 
\bar{u}\left(k',s'\right) 
\left(\frac{\epsilon^{\prime\ast}\cdot  
\gamma q'\cdot \gamma+2 \epsilon^{\prime\ast}\cdot k'} 
{2 k'\cdot q'}\gamma_\mu+\gamma_\mu  
\frac{q'\cdot\gamma \epsilon^{\prime\ast}\cdot \gamma 
-2\epsilon^{\prime\ast}\cdot k}{2k\cdot q'}\right) 
u\left(k,s\right)\nonumber\\ 
&&\times \bar{u}\left(k,s\right) 
\left(\gamma_\nu\frac{q'\cdot\gamma \epsilon^{\prime}\cdot \gamma 
+2\epsilon^{\prime}\cdot k'}{2k'\cdot q'} 
+\frac{\epsilon'\cdot \gamma q'\cdot \gamma 
-2 \epsilon'\cdot k}{2k\cdot q'}\gamma_\nu 
\right)u\left (k',s'\right). 
\end{eqnarray} 
   Since $\overline{l_{\mu\nu}^{\rm BH}}$ is a product of a term containing 
the final-state photon polarization vector times another term containing 
its complex conjugate, the sum over the polarizations can be carried out  
immediately yielding the well-known result 
\begin{equation} 
\label{polsum} 
\sum_{\lambda'}\epsilon^{\prime\ast}_\mu(\lambda') 
\epsilon^{\prime}_\nu(\lambda') 
=-g_{\mu\nu}. 
\end{equation} 
   This simple version holds true only for the case of conserved quantities, 
i.e., quantities which vanish after replacement of the polarization vector 
with the corresponding photon momentum.\footnote{For the general case, one  
would have to to use the  
complete formula given, e.g., in Ref.\ \cite{ms}.} 
   This is the case in the present situation.  
   Contracting the electron spinors and using trace techniques 
according to Ref.\ \cite{bdrel}, $\overline{l_{\mu\nu}^{\rm BH}}$  
can be expressed in terms of the four-momenta of the initial and final  
electrons  $k$ and $k'$, respectively, and the final-state photon  
four-momentum $q'$.  
   Finally, using on-shell conditions for the electron,  
$k^2=k^{\prime~2}=m_e^2$, we obtain  
\begin{eqnarray} 
\label{lbh2} 
\overline{l_{\mu\nu}^{\rm BH}}&=& 
-\frac{1}{2m_e^2k\cdot q'}\left(2 k_\mu k_\nu - q'_\mu k'_\nu - k'_\mu 
  q'_\nu+h_{\mu\nu}+k\cdot k' g_{\mu\nu}\right)  
\nonumber\\ && 
+ \frac{1}{2m_e^2k'\cdot q'} 
\left(2k'_\mu k'_\nu+q'_\mu k_\nu + k_\mu q'_\nu+h_{\mu\nu}+ k\cdot k' 
  g_{\mu\nu}\right)  
\nonumber \\ && 
-\frac{1}{2m_e^2 k\cdot q' k'\cdot q'} 
\left[k\cdot k'\left(q'_\mu k'_\nu +k'_\mu q'_\nu -q'_\mu k_\nu-k_\mu q'_\nu 
-2h_{\mu\nu}-2m_e^2 g_{\mu\nu}\right)+2m_e^2 q'_\mu q'_\nu\right] 
\nonumber \\ && 
-\frac{1}{m_e^2}g_{\mu\nu} 
-\frac{1}{2\left(k'\cdot q'\right)^2}\left(q'_\mu k_\nu + k_\mu 
  q'_\nu+h_{\mu\nu}+m_e^2 g_{\mu\nu}\right)  
\nonumber \\ && 
+\frac{1}{2\left(k\cdot q'\right)^2}\left(q'_\mu k'_\nu +k'_\mu 
  q'_\nu-h_{\mu\nu}-m_e^2 g_{\mu\nu}\right),  
 \end{eqnarray} 
where 
\begin{eqnarray} 
\label{hmunu} 
h_{\mu\nu}=k_\mu k'_\nu+k'_\mu k_\nu+\left(k'\cdot q'-k\cdot k'-k\cdot 
  q'\right)g_{\mu\nu}.    
\end{eqnarray} 
   Note that $\overline{l_{\mu\nu}^{\rm BH}}$ is symmetric under  
$k\leftrightarrow -k'$ as well as under $\mu\leftrightarrow\nu$. 
   Also $r^\mu\overline{l_{\mu\nu}^{\rm BH}}=0=r^\nu 
\overline{l_{\mu\nu}^{\rm BH}}$. 
   The analytical result for the contraction of Eq.\ (\ref{mbhsquared})  
is given in Eq.\ (\ref{mbh2}) of the Appendix.

   The VCS amplitude is obtained by contracting the pion VCS tensor 
with the initial virtual-photon and final real-photon polarization vectors  
$\epsilon_\mu=e\bar{u}\gamma_\mu u/q^2$ and  
$\epsilon^{\prime\ast}_\nu$, respectively, 
\begin{eqnarray} 
\label{mvcs} 
{\cal M}_{\rm VCS} &=& -\frac{i e^3}{q^2}\bar{u}\left(k',s'\right)\gamma_\mu 
u\left(k,s\right)\epsilon^{\prime\ast}_{\nu}{\cal M}^{\mu\nu}_{\rm VCS}. 
\end{eqnarray} 
   The generalized Born terms of the pion VCS tensor contain, in  
addition to the s-channel and u-channel pole terms, a ``contact term" in order 
to establish gauge invariance \cite{unkmeir00,lvov01,fearing98} 
\begin{equation} 
\label{mavcs} 
{\cal M}^{\mu\nu}_{\rm Born}= F\left(q^2\right)\left[\frac{(2 p_i+q)^\mu 
    (2p_f+q')^\nu}{s_2-m_{\pi}^2}+\frac{(2 p_f-q)^\mu 
    (2p_i-q')^\nu}{u_2-m_{\pi}^2}-2 g^{\mu\nu}\right], 
\end{equation} 
   where $u_2=(p_i-q')^2$.  
   The most general form of the non-Born tensor involves three independent 
structures with corresponding invariant amplitudes depending on three 
invariant variables (see Refs.\  
\cite{unkmeir00,lvov01,Drechsel_1997,Drechsel_1998}). 
   Here we will only consider the low-energy limit and make use of 
the ${\cal O}(p^4)$ predictions of Eq.\ (\ref{alphapp1}). 
   In this approximation the non-Born terms read 
\begin{equation} 
\label{mbvcs} 
{\cal M}^{\mu\nu}_{\rm NB}=-\frac{8\pi m_\pi}{e^2}  
\left(q^{\prime\mu} q^{\nu}-q \cdot q' g^{\mu\nu}\right)\alpha_L(q^2), 
\end{equation} 
   where $\alpha_L(q^2)$ is given in Eq.\ (\ref{alphapp1}). 
    
   The squared pure VCS matrix element involves the standard lepton tensor  
known from the one-photon exchange approximation in electro-production  
processes, 
\begin{eqnarray} 
\label{etamunu} 
\overline{\eta_{\mu\nu}}  
&=& \frac{1}{2}\sum_{s,s'}[\bar{u}(k',s')\gamma_\mu u(k,s)] 
[\bar{u}(k',s')\gamma_\nu u(k,s)]^\ast\nonumber\\ 
&=& \frac{1}{2m_e^2} 
\left[k'_\mu k_\nu+k_\mu k'_\nu+g_{\mu\nu}(m_e^2-k\cdot k')\right]\nonumber\\ 
&=&\frac{1}{4m_e^2}(K_\mu K_\nu-q_\mu q_\nu+ q^2 g_{\mu\nu}), 
\end{eqnarray} 
where $K=k+k'$ and $q=k-k'$.  
   The second equation is particularly useful when contracting with 
another  ``conserved'' tensor.  
   The lepton tensor is contracted  
with a hadronic tensor of the  
form\footnote{At ${\cal O}(p^4)$, for space-like 
$q^2$ all quantities are real, even for $s_2\geq (3m_\pi)^2$, and we could,  
in principle, omit the complex conjugation.  
However, for the sake of generality, 
we stick to the full notation.} 
\begin{equation} 
\label{squaredhadronicvcstensor} 
\overline{H^{\mu\nu}_{\rm VCS}}=\sum_{\lambda'}  
\epsilon^{\prime\ast}_\rho {\cal M}^{\mu\rho}_{\rm VCS}  
[\epsilon^{\prime\ast}_\sigma {\cal M}^{\nu\sigma}_{\rm VCS}]^\ast  
=-g_{\rho\sigma} {\cal M}^{\mu\rho}_{\rm VCS}  
{\cal M}^{\nu\sigma\ast}_{\rm VCS}, 
\end{equation} 
where, again, we made use of Eq.\ (\ref{polsum}). 
   Using Eq.\ (\ref{etamunu}) together with current conservation, we then 
obtain  
\begin{eqnarray} 
\label{squaredhadronicvcstensor2} 
\overline{|{\cal M}_{\rm VCS}|^2}&=&\left(\frac{e^3}{q^2}\right)^2  
\overline{\eta_{\mu\nu}}\,\overline{H^{\mu\nu}_{\rm VCS}}\nonumber\\ 
&=&-\left(\frac{e^3}{q^2}\right)^2 \frac{1}{4m_e^2}\left( 
K_\mu M^{\mu\rho}_{\rm VCS} K^\nu M^{{\rm VCS}\ast}_{\nu\rho} 
+q^2 M^{\mu\rho}_{\rm VCS} M^{{\rm VCS}\ast}_{\mu\rho}\right). 
\end{eqnarray} 
   Squaring the VCS tensor as in Eq.\ (\ref{squaredhadronicvcstensor2}) 
gives rise to three distinct contributions. 
   They involve, respectively, the contraction of the generalized Born  
terms, of the interference between the generalized Born terms and the residual
terms, and of the residual terms involving the polarizabilities. 
   The respective analytical results can be found in Eqs.\ (\ref{mborn2}), 
(\ref{mbornmnb}), and (\ref{mnb2}) of the Appendix.

   The interference term between the BH and the VCS amplitude can, again, be 
written as the contraction of a hadronic and a leptonic part.  
   Since the BH amplitude changes sign under the substitution 
$\pi^+\leftrightarrow\pi^-$, the interference term is sensitive to the pion 
charge.  
   The upper and lower signs apply to the cases of a $\pi^+$ and a $\pi^-$ 
beam, respectively, 
\begin{eqnarray} 
\label{BHVCSint} 
\overline{2 Re\{{\cal M}_{\rm VCS}^\ast{\cal M}_{\rm BH}\}} 
&=& 
\frac{1}{2}\sum_{s,s',\lambda'} 2 Re\{{\cal M}_{\rm VCS}^\ast 
{\cal M}_{\rm BH}\} 
\nonumber\\ 
&=&\pm 2 \frac{e^6}{r^2 q^2} F(r^2) (p_f+p_i)^\nu  
\overline{L_{\mu\nu\rho}^{\rm int}} Re\{M^{\rho\mu\ast}_{\rm VCS}\}, 
\end{eqnarray} 
where 
\begin{displaymath} 
\overline{L_{\mu\nu\rho}^{\rm int}}=\frac{1}{2}\sum_{s,s'} 
\bar{u}(k,s)\gamma_\rho u(k',s') L^{\rm BH}_{\mu\nu}. 
\end{displaymath} 
   Using trace techniques one obtains 
\begin{eqnarray} 
\label{lintmunurho} 
\overline{L_{\mu\nu\rho}^{\rm int}}&=&\frac{1}{2m_e^2}\left( 
\frac{1}{2k'\cdot q'}\bigg\{k'_\mu q'_\nu k_\rho 
+q'_\mu k'_\nu k_\rho 
+k_\mu q'_\nu k'_\rho +q'_\mu k_\nu k'_\rho 
-k_\mu k'_\nu q'_\rho + k'_\mu k_\nu q'_\rho\right.\nonumber\\ 
&&+g_{\mu\nu}[-(m_e^2-k\cdot k')q'_\rho-k'\cdot q' k_\rho-k\cdot q' k'_\rho] 
+g_{\mu\rho}[(m_e^2-k\cdot k')q'_\nu-k'\cdot q' k_\nu+k\cdot q' k'_\nu] 
\nonumber\\ 
&&+g_{\nu\rho}[(m_e^2-k\cdot k')q'_\mu+k'\cdot q' k_\mu-k\cdot q' k'_\mu]\bigg 
\} 
\left.+\frac{k'_\mu}{k'\cdot q'}\left[ 
k_\nu k'_\rho+k_\rho k'_\nu +(m_e^2-k\cdot k')g_{\nu\rho}\right]\right) 
\nonumber\\ 
&&-\frac{1}{2m_e^2}\Bigg( k\leftrightarrow -k'\Bigg). 
\end{eqnarray} 
   Finally, the interference term can be divided into the interference 
of the BH part with the VCS Born and VCS non-Born contributions, 
respectively. 
   The results are given in Eqs.\ (\ref{mbornmbh}) and (\ref{mnbmbh}) of the  
Appendix.

\section{General features of the VCS differential cross section} 
   Some general features of the fourfold differential cross section, 
Eq.\ (\ref{eq:xsection}), can be inferred from  
Eqs.\ (\ref{kin1}) - (\ref{mnbmbh}) of the Appendix. 
   The $s_1$ dependence is dominated by the $(s_1-m^2_e)^{-1}$ pole 
of the BH amplitude and the cross section varies according to this law 
with only a slight modification through the $s_1$ dependence of the VCS 
amplitude.  
   The $s_2$ dependence is dominated by the $(s_2-m^2_{\pi})^{-1}$ pole of  
the VCS amplitude with modifications due to the $s_2$ dependence of the BH  
amplitude, but in this case the modification is not so small as in the case  
of $s_1$.  
 
   The $r^2$ dependence is determined by the $1/(r^2)^2$ pole of the  
squared BH amplitude, and the $q^2$ dependence follows the  
$1/(q^2)^2$ pole behavior typical of electron scattering.  
   The energy of the outgoing pion is expected to be high 
and the angle is expected to be small according to $1/(r^2)^2$ behavior of the 
cross section.  
   The energy of the final-state photon is mainly expected to 
be low, as a consequence of the infrared divergence of the BH amplitude.  
   The angular behavior of the final photon is determined by the 
$1/(s_1-m^2_e)$ and $1/(s_2-m^2_{\pi})$ poles of the BH and VCS 
amplitudes, respectively. 
   The more interesting photons related to the GPs are 
expected to have higher energies.  
   The behavior of the final electron is completely determined by  
the $1/(q^2)^2$ behavior of the cross section. 
 
   The different behavior under the substitution  
$\pi^-\rightarrow\pi^+$ of ${\cal M}_{\rm BH}$ and ${\cal M}_{\rm VCS}$, 
\begin{eqnarray} 
\label{matr} 
{\cal M}_{\rm BH}(\pi^-)=-{\cal M}_{\rm BH}(\pi^+), \nonumber \\ 
{\cal M}_{\rm VCS}(\pi^-)={\cal M}_{\rm VCS}(\pi^+), 
\end{eqnarray} 
may be of use in identifying the contributions due to internal 
structure by comparing the reactions involving $\pi^-$ and $\pi^+$ 
beams for the same kinematics \cite{scherer}: 
\begin{equation} 
\label{subs} 
d\sigma(\pi^+)-d\sigma(\pi^-)\sim{4}\overline{Re\{{\cal M}_{\rm BH}(\pi^+) 
{\cal M}^*_{\rm VCS}(\pi^+)\}}. 
\end{equation} 
   This circumstance suggests the possibility of using such a subtraction to 
remove from the data sample the pure BH and VCS contributions.

\section{VCS event generator} 
\label{sec:generator} 
 
   The VCS event generator has been written based on the differential 
cross section of Eq.\ (\ref{eq:xsection}) in combination with the 
matrix elements of Eqs.\ (\ref{mbh2})--(\ref{mnbmbh}),  
using three-body-final-state kinematics.  
   The acceptance-rejection method \cite{ARM} is used for event generation.  
 
\subsection{Generation of the invariants} 
   The generation starts with the calculation of the squared invariant mass 
$s$ associated with the incoming pion beam and the electron target 
[see Eq.\ (\ref{ivar})].  
   In terms of laboratory (lab) variables, $s$ is given by 
\begin{equation} 
\label{calcs} 
s=m^2_{\pi}+m^2_e+2E_im_e, 
\end{equation} 
where $E_i$ is the incident beam energy. 
 
   The VCS cross section increases rapidly if either the direction of the 
outgoing real photon is close to the direction of one of the outgoing 
charged particles (due to the $1/(s_1-m^2_e)$ and $1/(s_2-m^2_{\pi})$ 
poles in the BH and VCS amplitudes, respectively),  
or if the energy of the outgoing real photon approaches zero  
(due to the infrared divergence of the $1/(s_1-m^2_e-q^2+r^2)$ BH pole).

   Therefore, if events are generated in the pole region, the efficiency of  
the acceptance-rejection method for the more interesting non-pole regions 
can be rather low.  
    In order to generate interesting events at an acceptable rate,  
the pole regions are cut.  
    Invariants are generated in the following regions: 
\begin{eqnarray} 
\label{s1s2r} 
1000\,m^2_e\,\leq & s_1 &\leq\,m^2_{\rho}, \nonumber \\ 
2\, m_{\pi}^2\,\leq & s_2 & \leq\,m^2_{\rho}, \nonumber \\ 
q^2_{\rm min}\,<& q^2 & <\, 2m_e^2-2m_e E_{k'}^{\rm min},\nonumber\\ 
r^2_{\rm min}\,<& r^2 & <\, -2m_e E_{q'}^{\rm min}+q^2+s_1-m^2_e. 
\end{eqnarray} 
 
\begin{figure}[ht] 
\vspace{1em}
\epsfxsize=16cm
\centerline{\epsffile{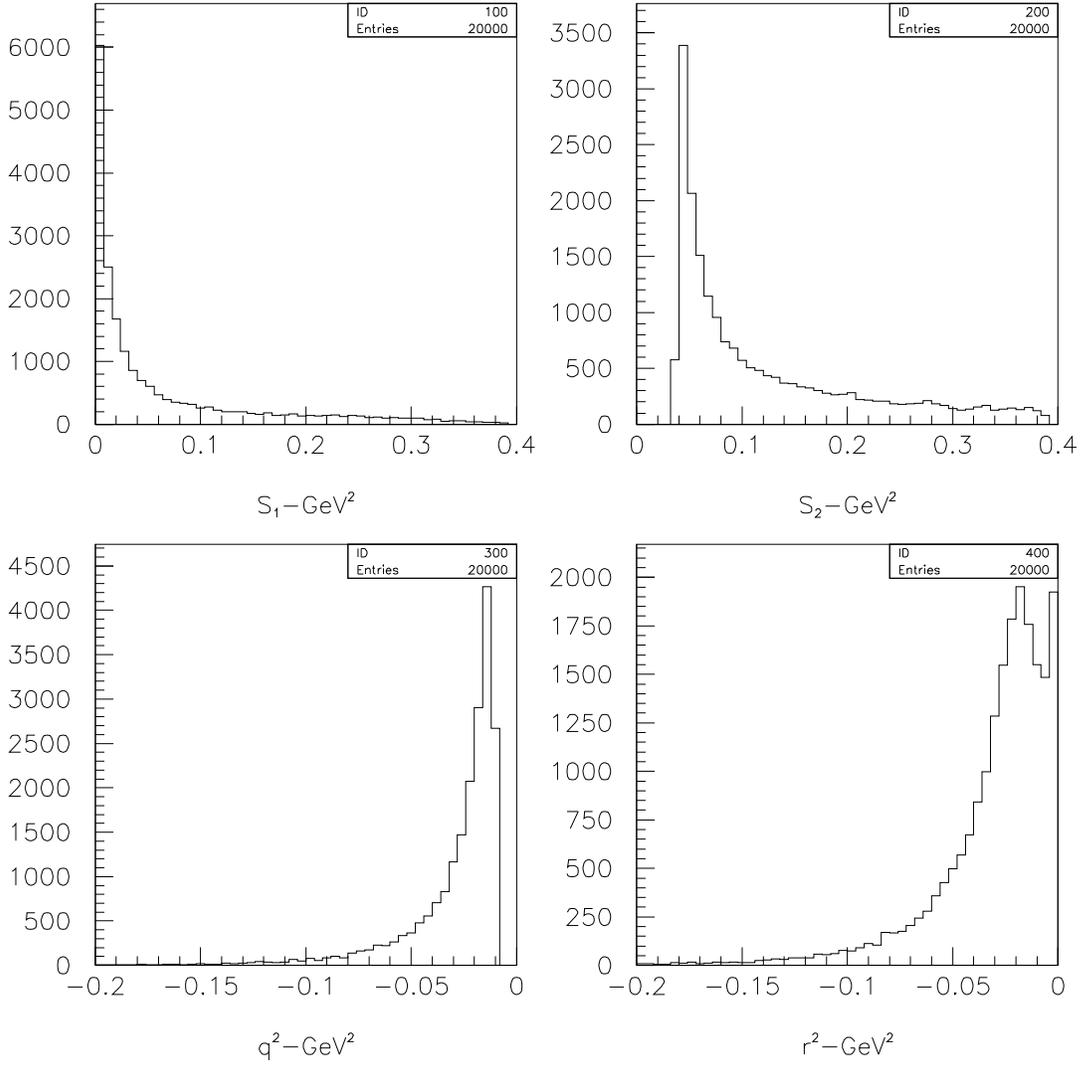}}
\vspace{1em}
\caption[]{Generated distributions of events plotted with 
      respect to the invariants of Eq.\ (\ref{ivar}).  
      The generation was done with a $\pi^-$ beam momentum of 650 GeV/c and  
      $\bar{\alpha}=6.8\times 10^{-43}$ cm$^3$.\label{fig:geninv}} 
\end{figure}

   We choose $E_{k'}^{\rm min}=10$ GeV in order to 
cut the 1/$q^2$ pole and to understand the trigger acceptance.  
   For the minimal photon energy we choose $E_{q'}^{\rm min}=5$ GeV in order 
to cut the infrared divergence of the BH amplitude.  
    For the generated invariants $s_1$, $s_2$, $q^2$, and $r^2$, the  
positivity of $-\Delta_4$ of Eq.\ (\ref{eq:delta}) is checked.  
   In case of a positive $\Delta_4$, the generated event is rejected and a  
new event is generated.  
   Actually, the requirement of a negative $\Delta_4$ is equivalent to the 
requirement of energy-momentum conservation.  
    For an accepted event, both the cross section of Eq.\ (\ref{eq:xsection})  
and the kinematical parameters of the outgoing particles are calculated.  
   In Fig.\ \ref{fig:geninv}, we show the generated distributions of events  
plotted with respect to the invariants of Eq.\ (\ref{ivar}).

\subsection{Calculation of energies and scattering angles of the outgoing 
        particles} 
\begin{figure}[ht] 
\vspace{1em}
\epsfxsize=16cm
\centerline{\epsffile{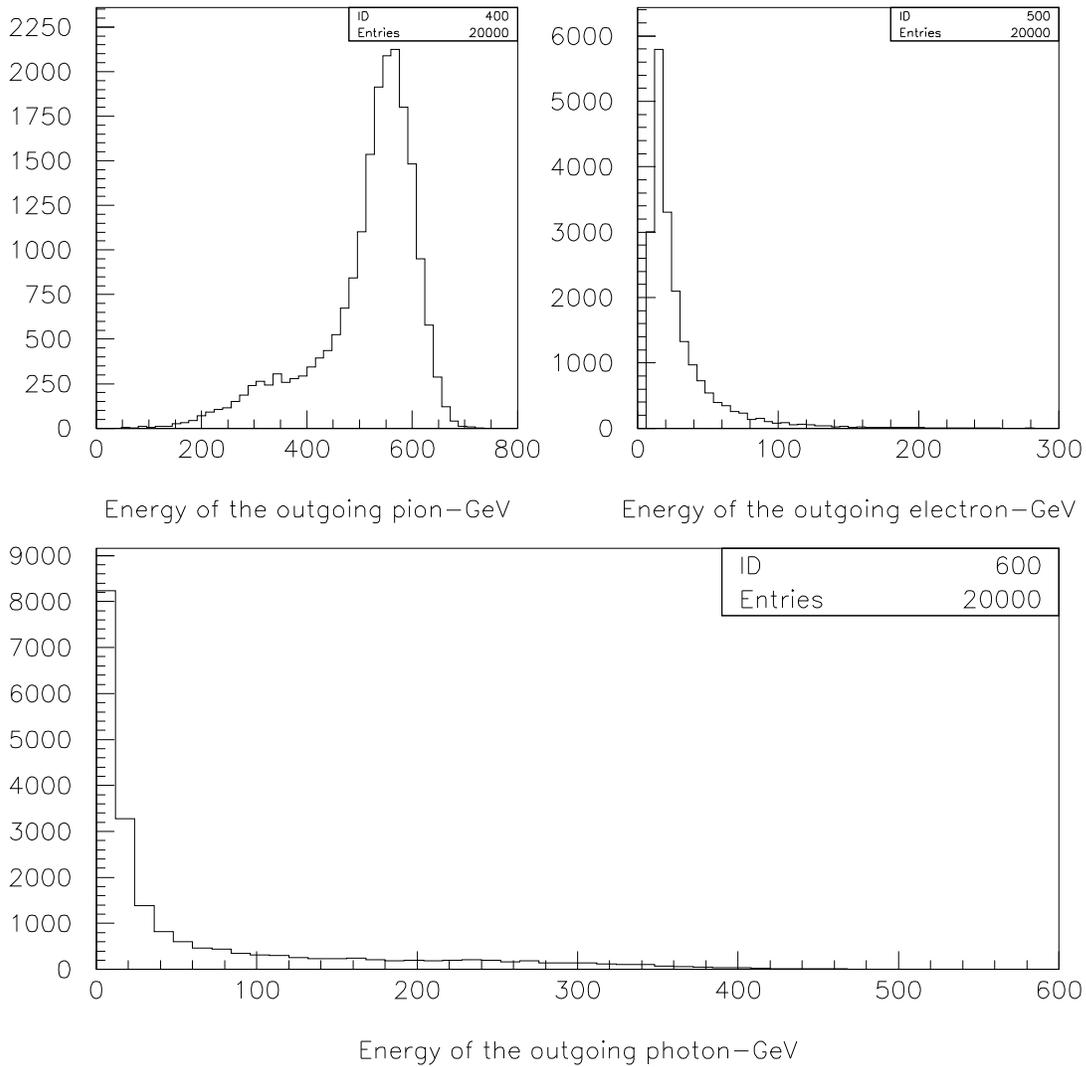}}
\vspace{1em}
\caption[]{Generated distributions of the energies of the 
      final-state particles. The generation was done with a $\pi^-$ beam 
      momentum of 650 GeV/c and $\bar{\alpha}=6.8\times 10^{-43}$ cm$^3$.
\label{fig:geenergy}}
\end{figure}

\begin{figure}[ht] 
\vspace{1em}
\epsfxsize=16cm
\centerline{\epsffile{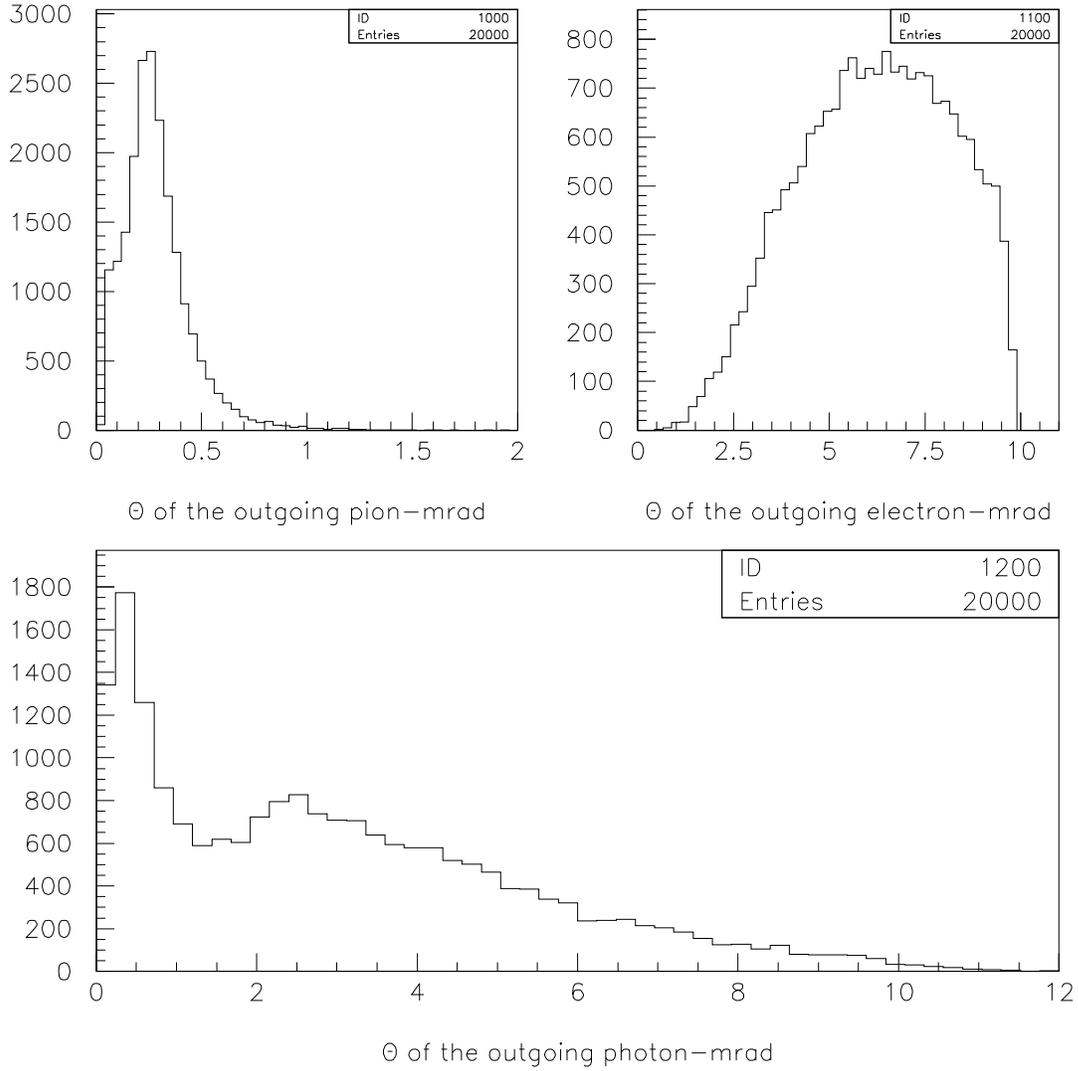}}
\vspace{1em}
\caption[]{Generated distributions of the $\theta$ scattering 
      angles of the final-state particles. The generation was done with a  
      $\pi^-$ beam momentum of 650 GeV/c and  
      $\bar{\alpha}=6.8\times 10^{-43}$ cm$^3$.\label{fig:geth}} 
\end{figure}

   Using the definitions of Eq.\ (\ref{ivar}) for the invariants, one obtains  
the following expressions for the energies and scattering angles of the  
outgoing particles with respect to the beam axis in terms of the lab 
variables: 
\begin{eqnarray} 
  \label{oute} 
  E_{k'}&=&\frac{2m_e^2-q^2}{2m_e}, \nonumber \\ 
  E_f&=&\frac{s+r^2-s_1-m^2_\pi}{2m_e}, \\ 
  E_{q'}&=&\frac{s_1+q^2-r^2-m^2_e}{2m_e}, \nonumber 
\end{eqnarray} 
and 
\begin{eqnarray} 
\label{outth} 
\cos(\theta_{k'})&=&\frac{s_2-s-q^2+m_e^2+2E_iE_{k'}}{ 
2|\vec{p}_i||\vec{p}_{k'}|},\nonumber \\ 
\cos(\theta_{p_f})&=&\frac{r^2-2m^2_{\pi}+2E_iE_f}{ 
2|\vec{p}_i||\vec{p}_f|},\\ 
\cos(\theta_{q'})&=&\frac{q^2-r^2-s_2+m^2_{\pi}+2 E_i E_{q'}}{ 
2|\vec{p}_i|E_{q'}}. \nonumber 
\end{eqnarray} 
The distributions generated for the energies and scattering angles are 
shown in Figs.\ \ref{fig:geenergy} and \ref{fig:geth}, respectively.

\subsection{Components of the 3-momenta of the outgoing particles} 
   We define the $(x,z)$ plane as the horizontal plane with the beam axis  
in the positive $z$ direction and the $y$ axis directed vertically, 
such that $\hat{e}_x$, $\hat{e}_y$, and $\hat{e}_z$ form a positively 
oriented Cartesian basis. 
   The spherical polar components of a generic 3-momentum  
$\vec{p}$ are defined as usual: 
\begin{eqnarray} 
\label{momcom} 
p_x&=&|\vec{p}\,|\sin(\theta)\cos(\phi), \nonumber \\ 
p_y&=&|\vec{p}\,|\sin(\theta)\sin(\phi), \nonumber \\ 
p_z&=&|\vec{p}\,|\cos(\theta). 
\end{eqnarray} 
   The $p_z$ components of the 3-momenta of the outgoing particles  
are calculated from the energies and scattering angles of the outgoing 
particles.  
   To calculate the $p_x$ and $p_y$ components, it is necessary to know the  
azimuthal angles which are obtained as follows. 
\begin{enumerate} 
\item From Eq.\ (\ref{ivar}),  
\begin{equation} 
s_1=m_e^2+2E_{k'}E_{q'}-2k'_x q'_x-2k'_y q'_y-2 k'_z q'_z, 
\end{equation} 
we obtain for the azimuthal angles of the final electron 
and photon  
\begin{equation} 
\label{eq:phi1} 
\frac{2E_{k'}E_{q'}+m_e^2-s_1}{2k'_z q'_z}= 
1+\tan(\theta_{k'})\tan(\theta_{q'})\cos(\phi_{k'}+\phi_{q'}). 
\end{equation} 
\item Analogously, from Eq.\ (\ref{ivar}),  
\begin{equation} 
s_2=m_{\pi}^2+2E_{f}E_{q'}-2{p_{f}}_x q'_x-2{p_{f}}_y q'_y -2{p_f}_z q'_z, 
\end{equation} 
we obtain for the azimuthal angles of the final pion 
and photon  
\begin{equation} 
\label{eq:phi2} 
\frac{2E_{p_f}E_{q'}+m_{\pi}^2-s_2}{2 {p_f}_z q'_z} 
= 1+\tan(\theta_{p_f})\tan(\theta_{q'})\cos(\phi_{p_f}+\phi_{q'}). 
\end{equation} 
\end{enumerate} 
We generate $\phi_{q'}$ randomly, and then use 
Eqs.\ (\ref{eq:phi1}) and (\ref{eq:phi2}) to determine $\phi_{p_f}$ and 
$\phi_{k'}$.  
 
\section{Sensitivity} 
\label{sec:sens} 
   In Ref.\ \cite{Pennington} and in the associated 
bibliography, detailed descriptions are given of the different 
theoretical models of the pion polarizabilities.  
   Here, we only collect the various predictions for the polarizabilities 
(see Table \ref{tab:predpol}). 
 
   For values of $\bar{\alpha}$ of the order of $10^{-42}$ cm$^3$  
(see Table \ref{tab:predpol}) one obtains for the ratio  
${\bar{\alpha}}/{(\frac{4}{3}\pi{r_{\pi}^3})}\sim{10^{-3}}$, 
where $r_{\pi}$ is the pion electromagnetic radius. 
   A comparison with the hydrogen atom, for which this ratio is of the order 
of 1, leads to the conclusion that the pion is a very rigid object, with 
strong interactions between its charged constituents. 
   As a consequence, we expect the effect of the polarizabilities on the  
RCS or VCS cross sections to be rather small. 
   Therefore a crucial component of the event generation is the  
determination of the VCS phase space region, where the data are  
expected to be sensitive to the pion polarizability.  
 
\begin{figure}[ht] 
\vspace{1em}
\epsfxsize=16cm
\centerline{\epsffile{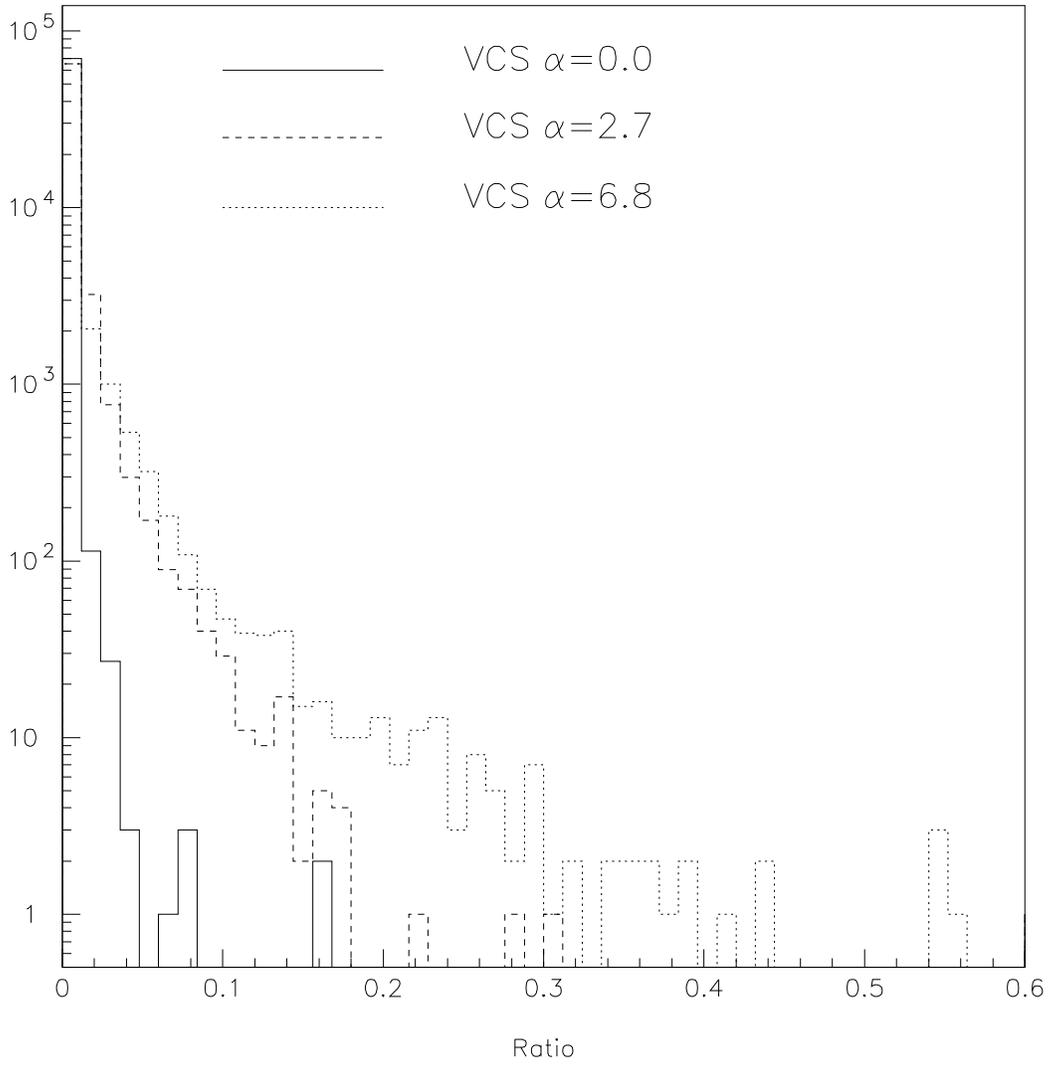}}
\vspace{1em}
\caption[]{The distributions of the  variable $Ratio$ 
      [see Eq.\ (\ref{ratio})] for different values of the pion 
      polarizability.\label{fig:ratio}} 
\end{figure}

\begin{figure}[ht] 
\vspace{1em}
\epsfxsize=16cm
\centerline{\epsffile{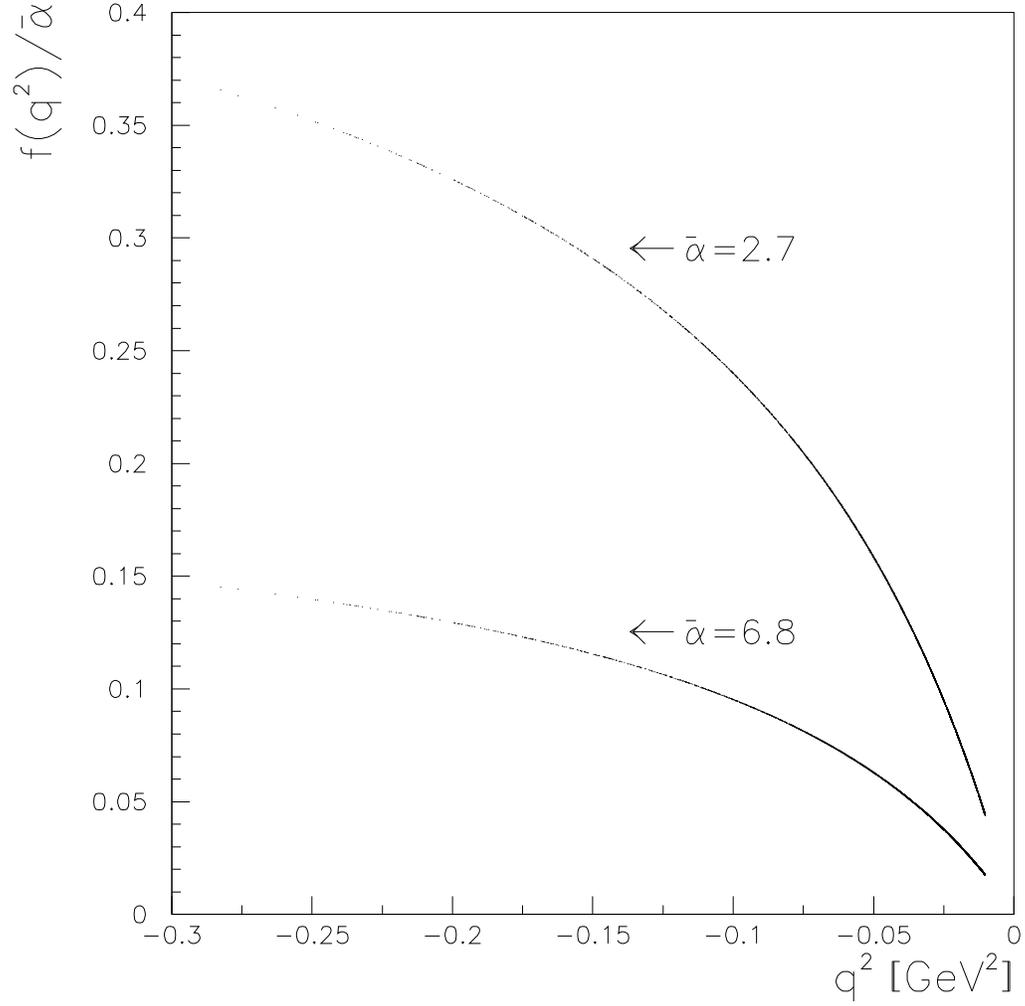}}
\vspace{1em}
\caption[]{Plot of $f(q^2)/(\bar{\alpha})$ 
      (Eq. (\ref{alphapp1})) versus $q^2$, for $\bar{\alpha}=2.7\times
      10^{-43}$ and $\bar{\alpha}=6.8\times 10^{-43}$.\label{fig:polratio}} 
\end{figure}

   To that end, we introduce the variable $Ratio$ as 
\begin{equation} 
\label{ratio} 
Ratio=\frac{2\overline{Re\{{\cal M}^\ast_{\rm Born} {\cal M}_{\rm NB}\}} 
+2 \overline{Re\{{\cal M}^\ast_{\rm NB} {\cal M}_{\rm BH}\}}+ 
\overline{|{\cal M}_{\rm NB}|^2}}{\overline{|{\cal M}|^2}}. 
\end{equation} 
   Here, $\overline{|{\cal M}|^2}$ is the total squared amplitude, 
whereas both $\overline{Re\{{\cal M}^\ast_{\rm Born} {\cal M}_{\rm NB}\}}$ and 
$\overline{Re\{{\cal M}^\ast_{\rm NB} {\cal M}_{\rm BH}\}}$ depend linearly 
and $\overline{|{\cal M}_{\rm NB}|^2}$ quadratically on the generalized 
dipole polarizability $\alpha_L(q^2)$ [see Eqs.\ (\ref{alphapp1}),
(\ref{mbornmnb}), (\ref{mnb2}), and (\ref{mnbmbh})]. 
   We generated VCS events for different values of $\bar{\alpha}$ in
   Eq. (\ref{alphapp1}).  
   The corresponding distributions of these events with respect to 
the variable $Ratio$ are shown in Fig.\ \ref{fig:ratio}. 
For $\bar{\alpha}=0$, we find   
$Ratio$ values less than 0.05, due to the $f(q^2)$ term 
$\alpha_L(q^2)$ [see the definition of Eq.\ (\ref{alphapp1})].
For $\bar{\alpha}=2.7\times 10^{-43}$ cm$^3$ and   
$\bar{\alpha}=6.8\times 10^{-43}$ cm$^3$ the  
``sensitivity'' reaches 20 \% and 50 \%, respectively.  
   The $q^2$-dependent term $f(q^2)$ in $\alpha_L(q^2)$ decreases relatively 
with increasing $\bar{\alpha}$, as shown in Fig.\ \ref{fig:polratio}.  

   When 
\begin{equation} 
\label{ratiocut} 
|Ratio|>0.05, 
\end{equation} 
the pion polarizability component of the VCS cross section exceeds 
5 \%.  
   The simulation shows that this occurs only for  
$\sim$0.1 \%, $\sim$0.6 \%, and $\sim$1.3 \% of the 
generated events for $\bar{\alpha}=0$, $\bar{\alpha}=2.7$, and  
$\bar{\alpha}=6.8\times 10^{-43}$ cm$^3$, respectively. 
 
\begin{figure}[ht] 
\vspace{1em}
\epsfxsize=16cm
\centerline{\epsffile{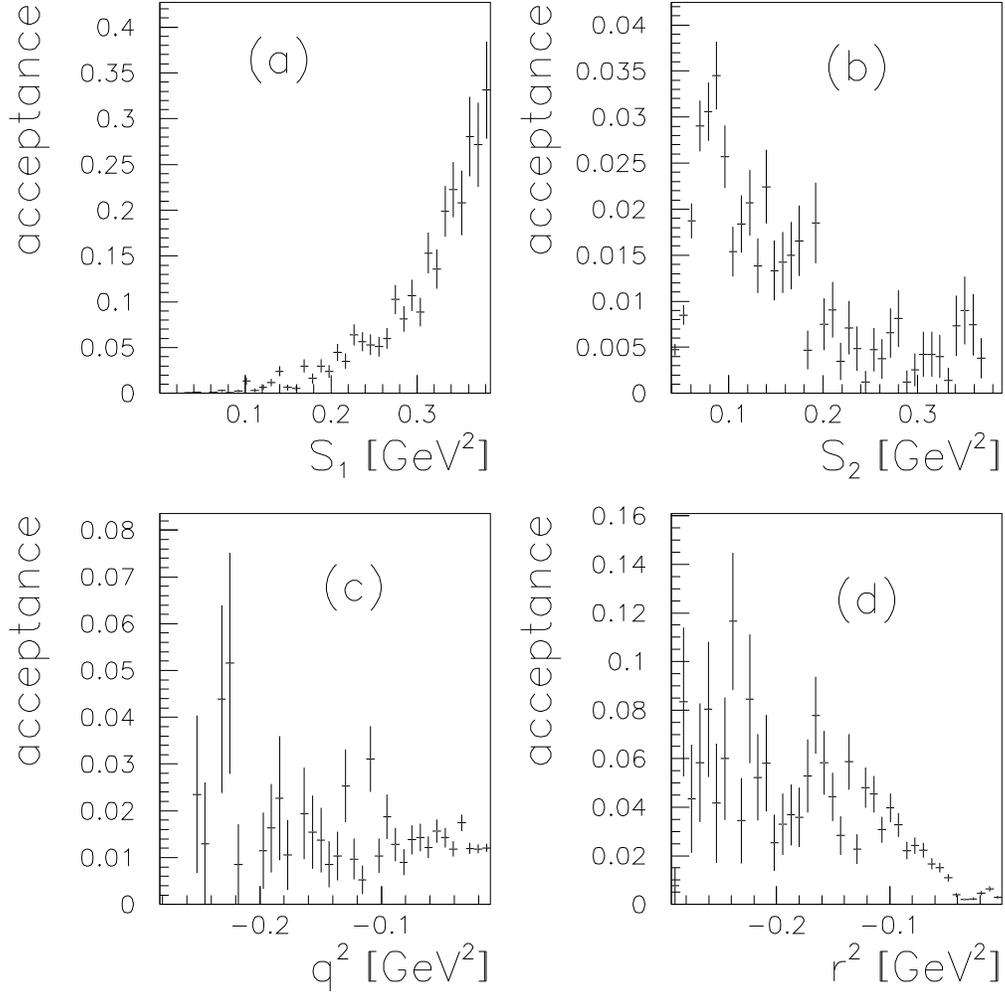}}
\vspace{1em}
\caption[]{The differential acceptance function of events that 
      satisfy the condition of Eq.\ (\ref{ratiocut}) for the 
      invariants for $\bar{\alpha}=6.8\times 10^{-43}$ cm$^3$.
\label{fig:invaracc}}
\end{figure}

\begin{figure}[ht] 
\vspace{1em}
\epsfxsize=16cm
\centerline{\epsffile{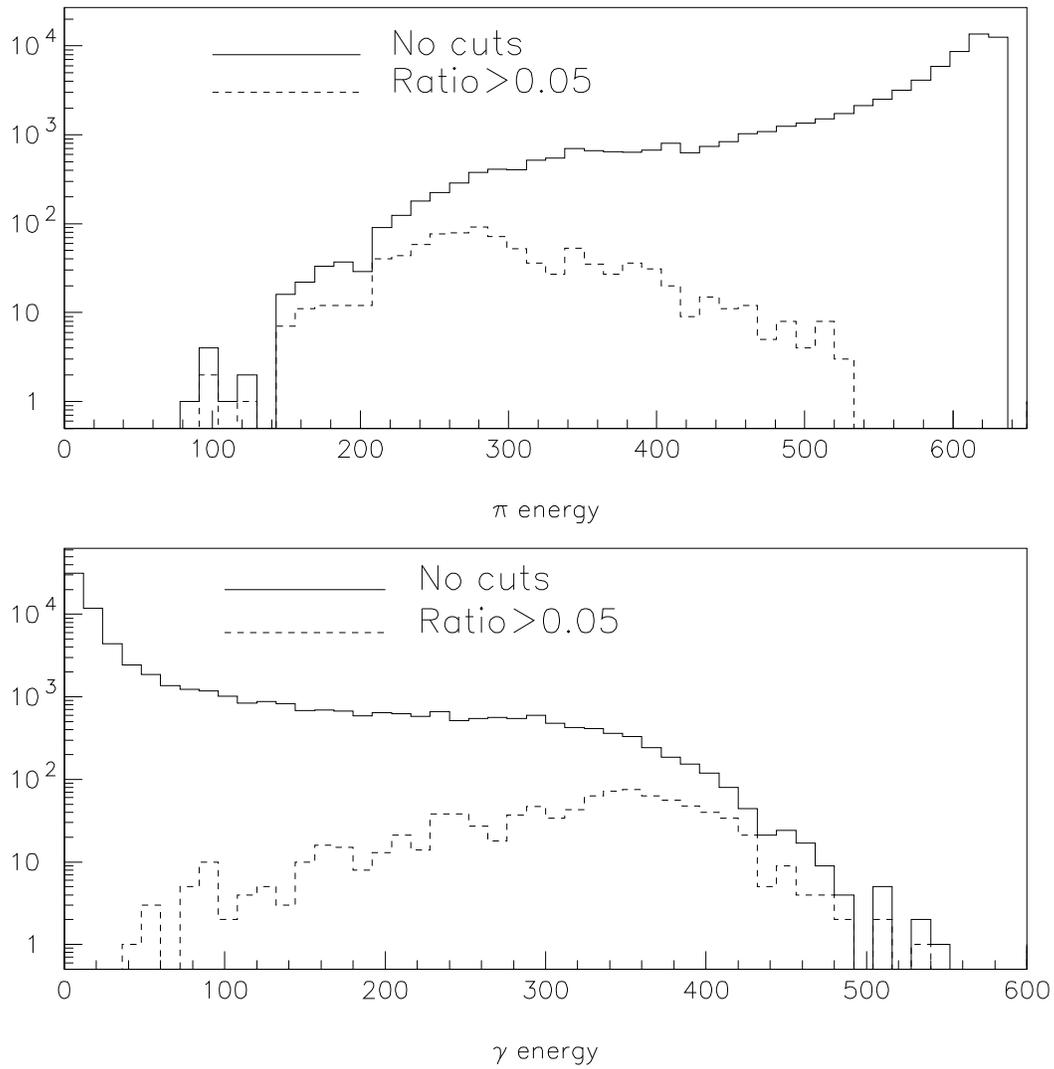}}
\vspace{1em}
\caption[]{The generated distribution of the energy of the pion (top) 
             and the photon (bottom). The dashed line shows 
      the part of the distribution with $Ratio>0.05$  
      [see Eq. (\ref{ratiocut})].\label{fig:energyacc}} 
\end{figure}

   The differential acceptance of these polarizability-sensitive events  
is shown as a function of invariants in Fig.\ \ref{fig:invaracc}, 
from which one finds that the sensitivity of VCS to 
pion structure is significant in the following cases:  
\begin{enumerate} 
\item the invariant mass $s_1$ is larger than 0.1 GeV$^2$, 
\item the absolute value of the squared momentum transfer $r^2$ is larger than 
      0.05 GeV$^2$,   
\item the invariant mass $s_2$ is in the range $0.05$ GeV$^2<s_2<0.2$ GeV$^2$. 
\end{enumerate} 
   Fig.\ \ref{fig:invaracc} (bottom left) shows that the differential 
acceptance function versus $q^2$ is approximately flat. Therefore, the 
dependence of the sensitivity to the electron energy is small.  
   Fig.\ \ref{fig:invaracc} (bottom right) shows that the sensitivity 
increases with increasing absolute value of the squared momentum 
transfer $r^2$. 
   Therefore,  the sensitivity is larger for lower values of the 
outgoing pion momentum [see Fig.\ \ref{fig:energyacc} (top)].  
   As a result of the infrared divergence of the BH amplitude,  
the sensitivity increases with increasing energy of the outgoing photon 
[see Fig.\ \ref{fig:energyacc} (bottom)].   
 
\begin{figure}[ht] 
\vspace{1em}
\epsfxsize=16cm
\centerline{\epsffile{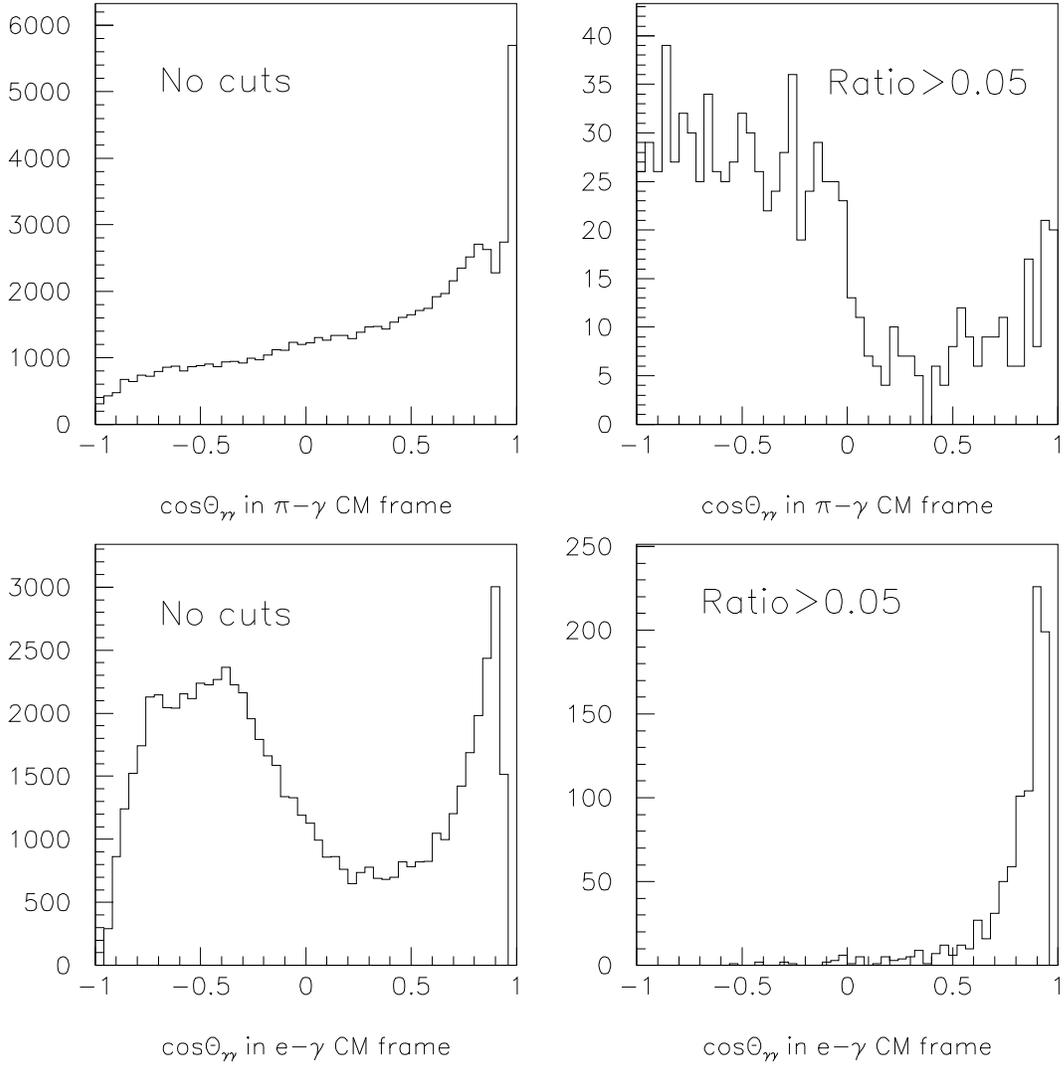}}
\vspace{1em}
\caption[]{The generated distribution of the angle $\theta^*_{\gamma\gamma}$ 
      in the $\pi\gamma$ (top) and $e\gamma$ (bottom) CM frames, respectively. 
      Distributions with $Ratio>0.05$ [see Eq.\ (\ref{ratiocut})] 
      are shown on the right.\label{fig:angacc}} 
\end{figure}

   Events with $Ratio>0.05$ are concentrated in the forward region of 
the angle $\theta^*_{\gamma\gamma}$ in the $e\gamma$ CM frame and in the 
backward region of the angle $\theta^*_{\gamma\gamma}$ of the $\pi\gamma$ CM 
frame, as shown in Fig.\ \ref{fig:angacc}.

\begin{figure}[ht] 
\vspace{1em}
\epsfxsize=16cm
\centerline{\epsffile{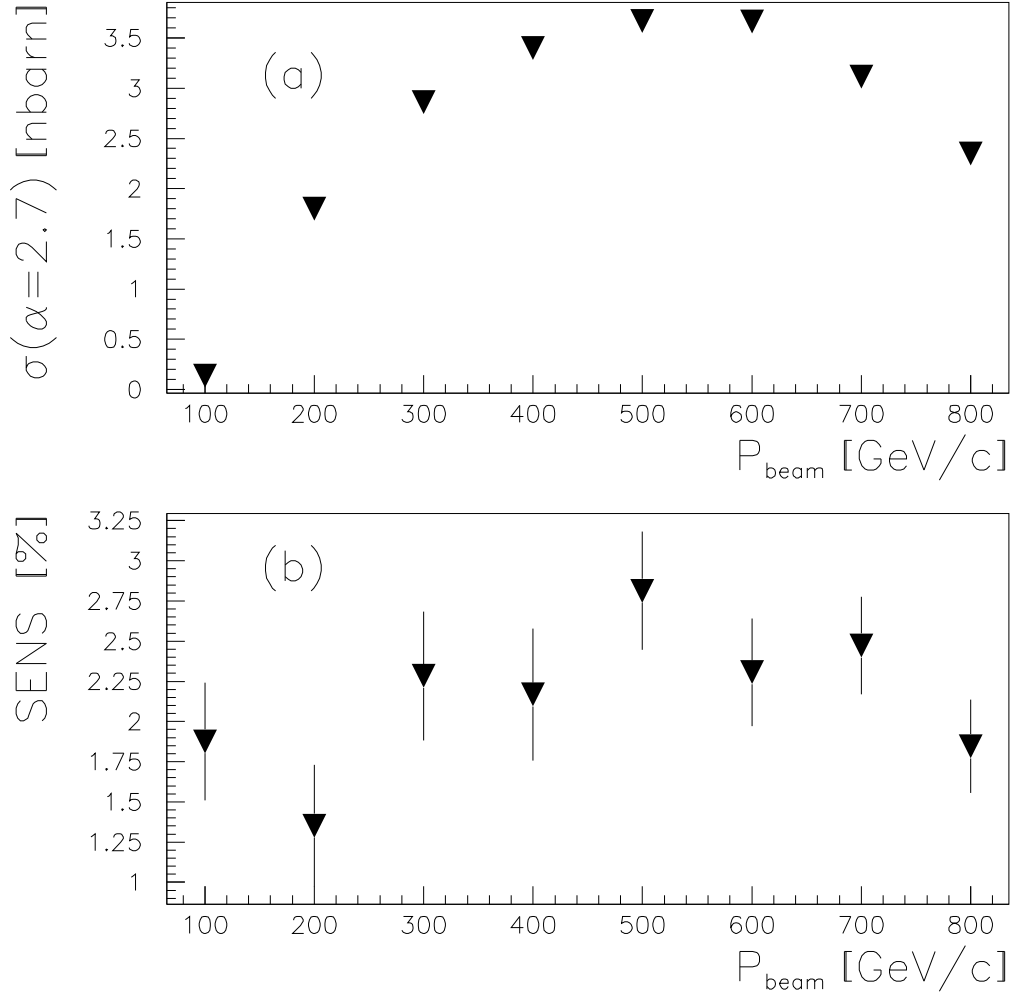}}
\vspace{1em}
\caption[]{(a) Total cross section of pion VCS for 
      $\bar{\alpha}=2.7 \times 10^{-43}$ cm$^3$ as a function of the  
      beam momentum including the requirements of Eq.\ (\ref{pigamtrig}); 
      (b) $SENS$ distribution as a function of the beam momentum including 
      the requirements of Eq. (\ref{pigamtrig}).\label{fig:piongamatrig}} 
\end{figure}

   As shown by the calculations, by implementing the requirements  
\begin{eqnarray} 
\label{pigamtrig} 
E_f<0.62E_i,\nonumber\\   
E_{q'}>0.15E_i, 
\end{eqnarray} 
one achieves 
\begin{equation} 
\label{sensnew} 
SENS 
=\frac{\sigma_{\rm VCS}(\bar{\alpha}=2.7)-\sigma_{\rm VCS}(\bar{\alpha}=6.8)} 
{\sigma_{\rm VCS}(\bar{\alpha}=2.7)}\sim{2.25}\%, 
\end{equation} 
as shown in Fig.\ \ref{fig:piongamatrig} (b).   
 
\begin{figure}[ht] 
\vspace{1em}
\epsfxsize=16cm
\centerline{\epsffile{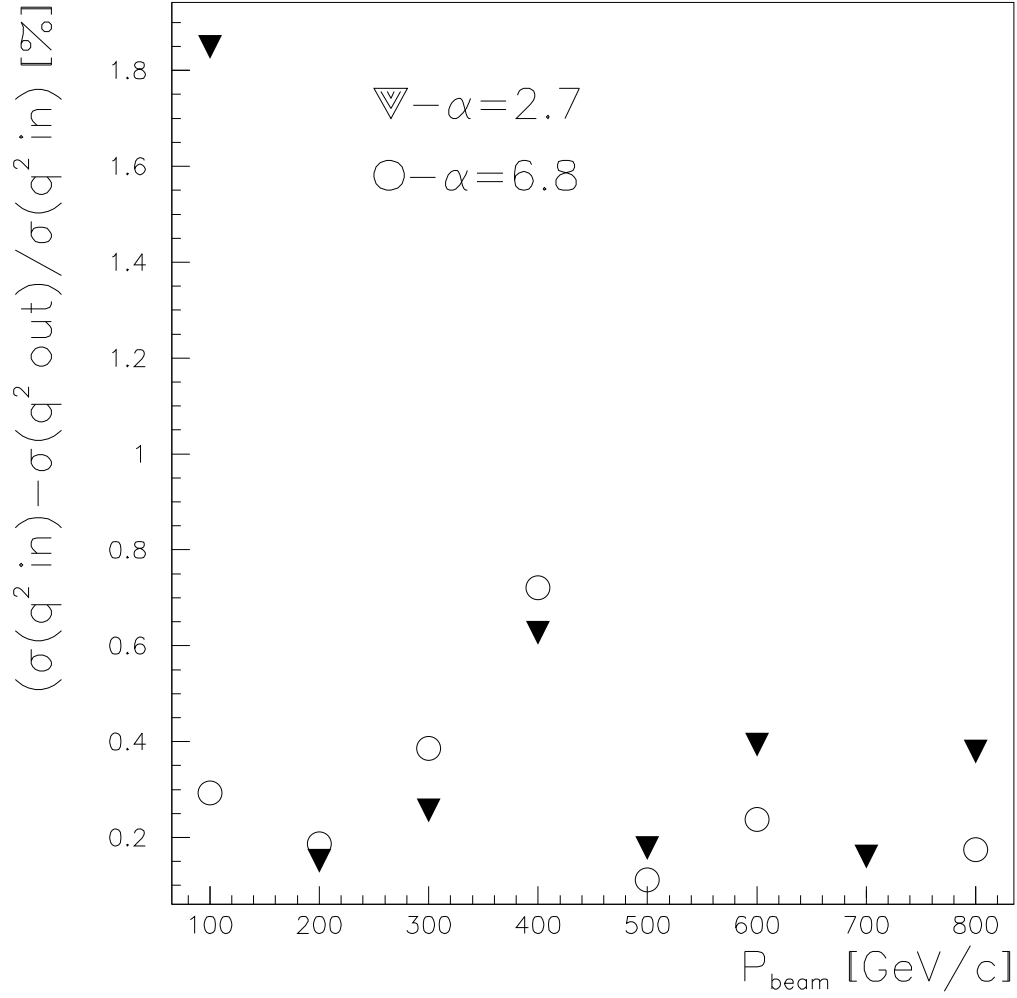}}
\vspace{1em}
\caption[]{Difference in the total (sensitive) cross section as a function  
           of the beam momentum with and without $q^2$ dependence of  
           ${\cal M}_{\rm NB}$ amplitude [see Eq.\ (\ref{mbvcs})] 
           for $\bar{\alpha}=2.7$ and $\bar{\alpha}=6.8 \times 10^{-43}$ 
           cm$^3$.\label{fig:senstoq2}} 
\end{figure}

   However, even when there is sensitivity to $\bar{\alpha}$, the 
sensitivity to the $q^2$ dependence is very low.  
   Fig.\ \ref{fig:senstoq2} shows the difference in the total 
(sensitive) cross section with and without $q^2$ dependence of the  
${\cal M}_{\rm NB}$ amplitude Eq. (\ref{mbvcs}) as a function of the beam 
momentum.  
    The percentage difference is only of the order of 0.3 \%. 
 
\section{VCS total cross section} 
\label{sec:totalcs} 
\begin{figure}[ht] 
\vspace{1em}
\epsfxsize=16cm
\centerline{\epsffile{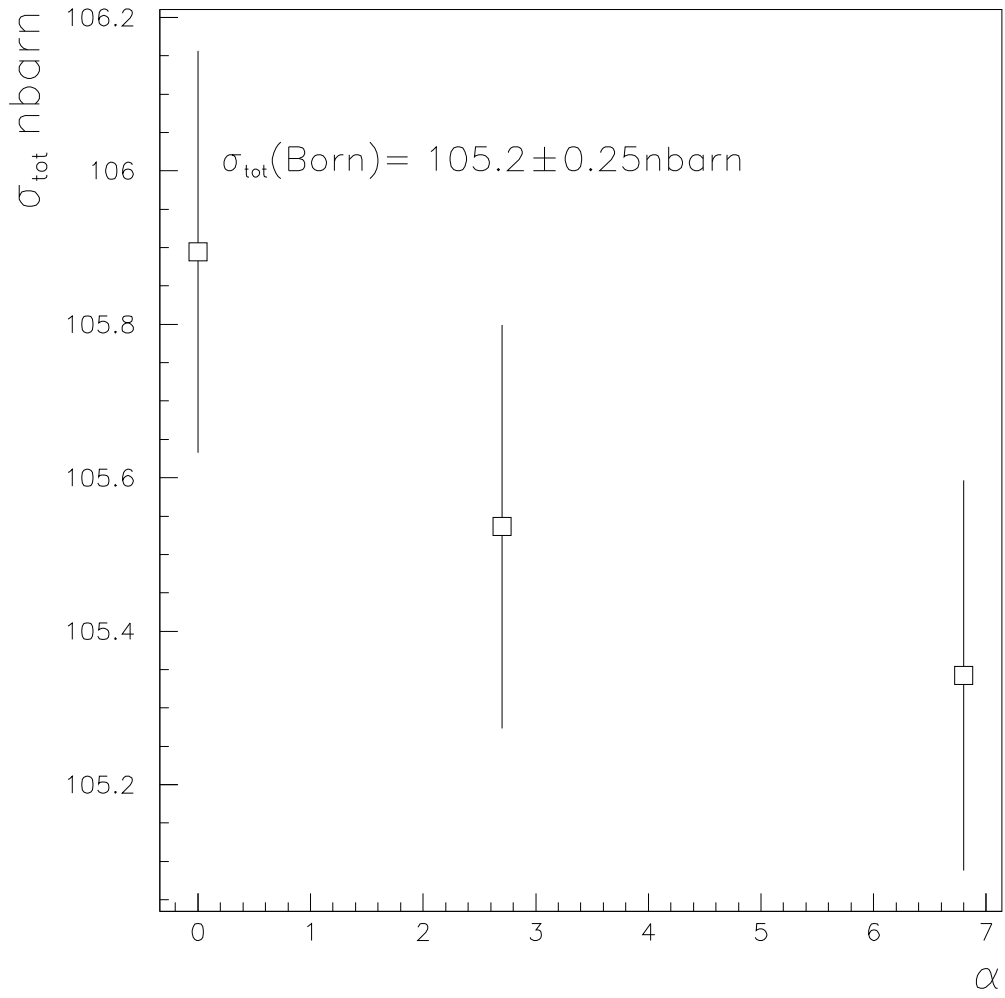}}
\vspace{1em}
\caption[]{Total cross section of $\pi^- e\to\pi^- e \gamma$  
             as a function of $\bar{\alpha}$. $\bar{\alpha}$ is given 
             in units of $10^{-43}$ cm$^3$.\label{fig:stots}} 
\end{figure}

   We have integrated the cross section of Eq.\ (\ref{eq:xsection})  
for the kinematic region specified in Eq.\ (\ref{s1s2r}) using two different  
Monte Carlo integration programs, namely, DIVON4 \cite{divon} and  
VEGAS \cite{vegas}. The results are 106.7$\pm$1.6 nbarn and
105.5$\pm$0.26 nbarn, respectively. For further calculations, we use
VEGAS, since the calculation errors given by VEGAS are about six times
smaller and the calculation is twice as fast as for DIVON4. Using
VEGAS, we calculate the Born cross sections of the reaction Eq.\
(\ref{react}) and the total cross section of the process Eq.\
(\ref{react}) for various values of $\bar{\alpha}$  including  
the $q^2$ dependence given by Eq.\ (\ref{mbvcs}). The results are
shown in Fig.\ \ref{fig:stots}, and the values are collected in Table
\ref{tab:stots}.      
 
   The small increase of the VCS total cross section for $\bar{\alpha}=0$, 
with respect to the VCS-Born cross section, is due to the $q^2$ 
dependent component of the ${\cal M}_{\rm NB} $ amplitude 
[see Eq.\ (\ref{mbvcs})]. The small decrease in the VCS total cross
section with increasing $\bar{\alpha}$ is due to the $\bar{\alpha}$
dependence of the ${\cal M}_{\rm NB}$ amplitude [see Eq.\ (\ref{mbvcs})].  

   We also calculated the $\pi$ VCS cross section for a second
kinematic region with 
\begin{equation}
\label{newq2}
-0.2\,\mbox{GeV}^2<q^2<-0.032\,\mbox{GeV}^2  
\end{equation}
for the $q^2$ range 
and ranges for $s_1$, $s_2$, and $r^2$ as given by Eq.\ (\ref{s1s2r}).
   This choice corresponds to the data region studied by
SELEX for Hadron-Electron elastic scattering \cite{ivo}. In this
region, using the VEGAS code, we find 34.7$\pm$0.1 nbarn for the $\pi$
VCS total cross section.

\section{Experimental status}  
Pion Virtual Compton Scattering (VCS) via the reaction  
$\pi{e}\rightarrow\pi e\gamma$ was observed in the Fermilab E781  
SELEX experiment \cite{exp}. SELEX used a 600 GeV $\pi^-$ beam    
incident on target atomic electrons, detecting the incident $\pi^-$ and  
the final state $\pi^-$, electron and $\gamma$. The number of  
reconstructed events and their distribution with   
respect to the kinematic variables (for the kinematic region studied)  
were shown to be in reasonable accord with the predictions.   
The limited statistics of the experiment did not allow deducing  
$\bar{\alpha}$. Ref. \cite{exp} discussed methods of improving the  
apparatus for a second-generation pion VCS experiment.

\section{Conclusion} 
 
   We have given explicit expressions for the differential cross section  
of the reaction $\pi^- e\to\pi^- e \gamma$, for which the invariant 
amplitude has been calculated at the one-loop level, ${\cal O}(p^4)$,  
in chiral perturbation theory.  
   The non-Born part of the VCS amplitude depends on the generalized 
dipole polarizability $\alpha_L(q^2)$. 
   In principle, there exist three generalized dipole polarizabilities 
which, however, are degenerate at the one-loop level. 
   The total cross section has been calculated using Monte Carlo integration 
programs. 
   With the VCS event generation we were able to identify regions of 
phase space which are sensitive to the pion polarizability. 
\acknowledgments 
   This work was supported by the Deutsche Forschungsgemeinschaft (SFB 201 
and SFB 443), and the Israel Science Foundation founded by the Israel 
Academy of Sciences and Humanities. 
 
\appendix 
\section{Analytical expressions of the squared invariant amplitudes} 
   In this appendix we list the analytical results for the squared  
invariant amplitudes as obtained after contraction of the relevant 
tensors. 
   The scalar products appearing naturally after contraction can 
be written in terms of the invariants of Eq.\ (\ref{ivar}) as 
\begin{eqnarray} 
\label{kin1} 
K^2 &=& (k+k')^2= 4m_e^2-q^2,\nonumber\\ 
P^2 &=& (p_i+p_f)^2 = 4m_{\pi}^2-r^2,\nonumber\\ 
q\cdot q'&=& \frac{1}{2}\left(q^2-r^2\right),\nonumber\\ 
P\cdot q &=& P\cdot q'=s_2-m_{\pi}^2-\frac{1}{2}(q^2-r^2),\nonumber \\ 
K\cdot q &=& (k+k')\cdot(k-k')=0,\nonumber\\ 
K\cdot q'&=& s_1-m_e^2+\frac{1}{2}(q^2-r^2),\nonumber\\ 
K\cdot P &=& 2s-s_1-s_2-m_{\pi}^2-m_e^2+\frac{1}{2}(q^2+r^2)\nonumber\\ 
2p_i\cdot K&=& 2s-2m_e^2-s_2+q^2-m_\pi^2\nonumber\\ 
2p_f \cdot K&=&2s-2s_1-m_\pi^2+r^2-s_2\nonumber\\ 
u_2-m_\pi^2&=&(p_i-q')^2-m_\pi^2=m_\pi^2+q^2-s_2-r^2.  
\end{eqnarray} 
   The squared contributions read: 
\begin{eqnarray} 
\label{mbh2} 
\overline{|{\cal M}_{\rm BH}|^2} &=& - \left[\frac{e^3 F(r^2)}{r^2}\right]^2 
\left[\frac{2}{(K\cdot q')^2-(q\cdot q')^2}\right]^2 \nonumber\\ 
&&\times \Bigg((K\cdot q')^2(P\cdot q')^2 
+(q\cdot q')^2[P^2 r^2 +(K \cdot P)^2] 
+ 2(K\cdot P) (P \cdot q')(K\cdot q')(q\cdot q')\nonumber\\ 
&&+\frac{(K\cdot q')^2-(q\cdot q')^2}{4 m_e^2} 
\left\{P^2\left[r^2 q^2+(K\cdot q')^2+(q\cdot q')^2\right] 
+r^2\left[(P \cdot q')^2+(K\cdot P)^2\right]\right\}\Bigg),\nonumber\\ 
&&\\ 
\label{mborn2} 
\overline{|{\cal M}_{\rm Born}|^2}&=& -\left[\frac{e^3 F(q^2)}{q^2}\right]^2 
\frac{1}{m_e^2} 
\Bigg\{ 
\frac{m_\pi^2 \left[q^2\left(P^2+s_2-u_2\right)+ 
(2p_i\cdot K)^2\right]}{\left(s_2-m_{\pi}^2\right)^2}\nonumber \\ && 
+\frac{m_\pi^2\left[q^2\left(P^2-s_2+u_2\right) 
+(2p_f\cdot K)^2\right]}{\left(u_2-m_{\pi}^2\right)^2}+2q^2+K^2\nonumber\\&& 
+\frac{\left(2p_i\cdot K\right)\left(2p_f\cdot K\right)+P^2 q^2} 
{\left(s_2-m_\pi^2\right)\left(u_2-m_\pi^2\right)}\left(2 m_\pi^2-q^2\right) 
\Bigg\}, 
\end{eqnarray} 
 
\begin{eqnarray} 
\label{mbornmnb} 
\overline{2 Re\{{\cal M}_{\rm Born}{\cal M}_{\rm NB}^\ast\}}&=& 
-\left(\frac{e^3}{q^2}\right)^2  
\left[\frac{8\pi m_\pi}{e^2}Re\{\alpha_L(q^2)\}F(q^2)\right] \frac{1}{m_e^2} 
\nonumber\\ 
&&\times 
\Bigg\{(p_i\cdot K p_f\cdot K+q^2 m_\pi^2) 
\frac{[q^2-r^2]^2}{(s_2-m_\pi^2)(u_2-m_\pi^2)}\nonumber\\ 
&&+p_i\cdot K q'\cdot K \frac{u_2-m_\pi^2-q^2}{s_2-m_\pi^2} 
-p_f\cdot K q'\cdot K \frac{s_2-m_\pi^2-q^2}{u_2-m_\pi^2}\nonumber\\ 
&&-\frac{1}{2}K^2(q^2-r^2)-(q^2)^2\Bigg\}, 
\end{eqnarray}

\begin{eqnarray} 
\label{mnb2} 
\overline{|{\cal M}_{\rm NB}|^2}&=& 
-\left(\frac{e^3}{q^2}\right)^2 \frac{1}{4 m_e^2} 
\left(\frac{8\pi m_\pi}{e^2}\right)^2 |\alpha_L(q^2)|^2 
[(q\cdot q')^2 K^2+q^2 (K\cdot q')^2+2 q^2 (q\cdot q')^2]. 
\end{eqnarray} 
 
For the evaluation of the Bethe-Heitler-VCS interference term we quote 
an intermediate result: 
\begin{eqnarray} 
\label{plint} 
2 P^\nu \overline{L^{\rm int}_{\mu\nu\rho}}&=& 
\frac{1}{m_e^2}\frac{1}{(K\cdot q')^2-(q\cdot q')^2} 
\Bigg\{ K_\rho K_\mu(P\cdot q' K\cdot q'+P\cdot K q\cdot q') 
-K_\rho P_\mu (K\cdot q')^2\nonumber\\ 
&&-K_\rho q_\mu P\cdot K K\cdot q' 
+P_\rho K_\mu q\cdot q'(q^2-q\cdot q') 
-P_\rho q_\mu K\cdot q'(q^2-q\cdot q')\nonumber\\ 
&&+q'_\rho K_\mu P\cdot q' q\cdot q' 
- q'_\rho P_\mu K\cdot q' q^2 
-q'_\rho q_\mu P\cdot K q\cdot q'\nonumber\\ 
&& 
+g_{\rho\mu}[(q^2-q\cdot q') P\cdot q' K\cdot q' 
+P\cdot K (q\cdot q')^2]\Bigg\}, 
\end{eqnarray} 
where we have rewritten $k=(K+q)/2$ and $k'=(K-q)/2$, and where 
terms proportional to $q_\rho$ and $q'_\mu$ have been dropped since 
they are contracted with the conserved VCS tensor. 
 
   The interference between the VCS Born terms and the BH terms is given 
by

\begin{eqnarray} 
\label{mbornmbh} 
\overline{2 Re\{{\cal M}_{\rm Born}^\ast{\cal M}_{\rm BH}\}} 
&=&\pm \frac{e^6}{q^2r^2}F(q^2)F(r^2)\frac{1}{m_{e}^2[\left(K\cdot q'\right)^2 
-\left(q\cdot q'\right)^2]} 
\nonumber\\ 
&&\times\left[\left(\frac{1}{s_2-m_\pi^2}+\frac{1}{u_2-m_\pi^2}\right)A+ 
\left(\frac{1}{s_2-m_\pi^2}-\frac{1}{u_2-m_\pi^2}\right)B 
+C\right], 
\end{eqnarray} 
where 
\begin{eqnarray*} 
A&=& P\cdot K q\cdot q'[(P\cdot K)^2+P^2 q^2]\\ 
&&-K\cdot q'\{(P^2+q^2)(K\cdot q' P\cdot K+q^2 P\cdot q') 
+P\cdot q'[(K\cdot q')^2 -2q^2 q\cdot q']\},\\ 
B&=&-K\cdot q'\{ (q^2-q\cdot q')[(P\cdot K)^2+P^2 q^2] 
+P^2 (K\cdot q')^2 +P\cdot q' P\cdot K K\cdot q'\},\\ 
C&=&-2[K\cdot q' P\cdot q'(2q^2-q\cdot q' +K^2) 
+P\cdot K q\cdot q'(q^2+q\cdot q'+K^2)-P\cdot K (K\cdot q')^2]. 
\end{eqnarray*} 
   In Eq.\ (\ref{mbornmbh}) the upper and lower signs refer to a $\pi^+$ and 
a $\pi^-$ beam, respectively.

   Finally, the interference between the non-Born terms of the VCS amplitude 
and the BH terms reads 
\begin{eqnarray} 
\label{mnbmbh} 
\overline{2 Re\{{\cal M}_{\rm NB}^\ast{\cal M}_{\rm BH}\}} 
&=& \pm 8\pi m_\pi Re\{\alpha_L(q^2)\}\frac{e^4}{q^2r^2}F(r^2) 
\frac{1}{m_{e}^2[(K\cdot q')^2-(q\cdot q')^2]}\nonumber\\ 
&&\times \Bigg\{P\cdot K[(K\cdot q')^2(q^2-q\cdot q')+(q\cdot q')^3 
+(K^2+q^2)(q\cdot q')^2]\nonumber\\ 
&&+P\cdot q' K\cdot q'[(K\cdot q')^2+(q^2-q\cdot q')(q^2+q\cdot q') 
+K^2 q\cdot q']\Bigg\}, 
\end{eqnarray} 
where the upper and lower signs refer to a $\pi^+$ and a $\pi^-$  
beam, respectively. 
    Here we made use of the approximation of Eq.\ (\ref{mbvcs}) which is valid 
at ${\cal O}(p^4)$ in ChPT.

\begin{table}[ht] 
\begin{center} 
\begin{tabular}{|c|c|c|} 
\hline 
Reaction& $\bar{\alpha}$ & Experiment \\ 
\hline 
$\pi^-Z\rightarrow \pi^-Z\gamma$ &$6.8\pm 1.4\pm 1.2$ & 
Serpukov \cite{Antipov} \\ 
$\gamma p \rightarrow \gamma \pi^+ n$ &$20\pm 12$ & Lebedev 
\cite{aibergenov}\\ 
$\gamma\gamma \rightarrow \pi^+\pi^-$ &$2.2\pm 1.6$ &MARK-2 \cite{Boyer}\\ 
\hline 
\end{tabular} 
\caption{Experimental values of $\bar{\alpha}$ in units of $10^{-43}$ 
cm$^3$.} 
\label{tab:mespol} 
\end{center} 
\end{table} 

\begin{table}[ht] 
\begin{center} 
\begin{tabular}{|c|c|c|} 
\hline 
Theoretical model&$\bar{\alpha}$&$\bar{\beta}$\\ 
\hline 
Chiral quark model \cite{Volkov}            &$8.0$        &$ -7.8$\\ 
Nonrelativistic quark model \cite{Schoeberl}&0.05         &-\\ 
NJL model \cite{Bernard}                    &$10.5 - 12.5$&$-(10.3 -11.8)$\\ 
ChPT [${\cal O}(p^4)$] \cite{overview,Babusci} &$2.68\pm 
0.42$&$-(2.68-2.61)\pm 0.42$\\ 
ChPT [${\cal O}(p^6)$] \cite{buergi96}       &$2.4\pm 
0.5$&$-2.1\pm 0.5$\\   
Quark confinement model \cite{Ivanov92}       &$3.63$&$-3.41$\\ 
QCD sum rules \cite{Lavelle94}                &$5.6\pm 0.5$&-\\ 
\hline 
\end{tabular} 
\caption{Theoretical predictions of the polarizabilities in 
  units of $10^{-43}$ cm$^3$.} 
\label{tab:predpol} 
\end{center} 
\end{table} 
 
\begin{table}[ht] 
\begin{center} 
\begin{tabular}{|c|c|c|} 
\hline 
&Total cross section (nb)&Error (nb)\\ 
\hline 
Born              &105.2&0.25\\ 
$\bar{\alpha}$=0.0&105.9&0.26\\ 
$\bar{\alpha}$=2.7&105.5&0.26\\ 
$\bar{\alpha}$=6.8&105.3&0.25\\ 
\hline 
\end{tabular} 
\caption{Born and total cross sections of the reaction Eq.\ (\ref{react}) 
for a negative pion as a function of $\bar{\alpha}$ ($\bar{\alpha}$ is 
given in units of $10^{-43}$ cm$^3$).} 
\label{tab:stots} 
\end{center} 
\end{table} 

\end{document}